\documentclass[journal]{IEEEtran}
\usepackage{algorithm}
\usepackage{algorithmic}
\usepackage{cite}
\usepackage{bbm}
\usepackage{amsmath}
\usepackage{amsfonts,amssymb}
\usepackage{graphicx}
\usepackage{colortbl}
\usepackage{adjustbox}

\allowdisplaybreaks[4]
\usepackage{enumitem}
\usepackage{booktabs}
\usepackage{xcolor,colortbl}
\definecolor{myc1}{RGB}{127 255 212}
\definecolor{myc2}{RGB}{255 181 197}
 \usepackage{bm}
 \usepackage{multirow}
\allowdisplaybreaks[4]
\usepackage{amsthm}
\usepackage{amssymb}
\usepackage{pgfplots}
\usepackage[utf8]{inputenc}
\usepackage{comment}
\definecolor{myblue}{RGB}{105,89,205}
\definecolor{myorange}{RGB}{238,92,66}
\definecolor{mygrey}{RGB}{205,201,201}

\begin{document}

\title{Global Optimality of Inverter \\Dynamic Voltage Support}
\author{
	Yifei~Guo,~\IEEEmembership{Member,~IEEE,}
	Bikash C. Pal,~\IEEEmembership{Fellow,~IEEE,}
	Rabih A. Jabr,~\IEEEmembership{Fellow,~IEEE}
\thanks{This work was supported by the EPSRC-NSFC Programme on Sustainable Energy Supply. (\emph{Corresponding author: Bikash C. Pal})}
\thanks{Yifei Guo and Bikash C. Pal are with the Electrical and Electronic
Engineering Department, Imperial College London, London SW7 2AZ, U.K. (e-mail: yifei.guo@imperial.ac.uk; b.pal@imperial.ac.uk)}
\thanks{Rabih A. Jabr is with the Department of Electrical and Computer
Engineering, American University of Beirut, Beirut 1107 2020, Lebanon
(e-mail: rabih.jabr@aub.edu.lb)}
}
\markboth{}%
{Shell \MakeLowercase{\textit{et al.}}: Bare Demo of IEEEtran.cls for IEEE Journals}
\maketitle

\begin{abstract}
This paper investigates the dynamic voltage support (DVS) control of inverter-based resources (IBRs) under voltage sags to enhance the low-voltage ride-through performance. We first revisit the prevalent droop control from an optimization perspective to elaborate on why it usually suffers from suboptimality. Then, we formulate the DVS problem as an optimization program that maximizes the positive-sequence voltage magnitude at the point of common coupling (PCC) subject to the current, active power, and stability constraints. The program is inherently nonconvex due to the active power limits, of which the global optimality is not guaranteed by off-the-shelf solvers. In this context, we perform the optimality analysis to explore the global optimum analytically. It is found that the unique global optimum has three scenarios/stages (S1--S3), which depends on the specific relationship among grid voltage, grid strength, as well as physical limits of IBRs. The closed-form solutions in S1 and S3 are derived and the optimality conditions for S2 are provided, which guarantees the optimality and compatibility with the fast real-time control. We implement the optimum with a grid-connected photovoltaic (PV) power plant by integrating a DVS controller. 
Dynamic simulations are carried out under different scenarios to test our proposal and compare it with other existing methods. Additionally, the robustness of optimality against model errors is discussed and numerically demonstrated.
\end{abstract}

\section{Introduction}
\IEEEPARstart{W}ith the increasing penetration of inverter-based resources (IBRs), the philosophy of disconnecting IBRs from the system ``at first sign of trouble" could exacerbate system instability under grid faults \cite{AG2009}. To avoid cascading trip-off events while enhancing system immunity against disturbances, the low-voltage ride-through requirements in grid codes have evolved from the quick disconnection to the dynamic voltage support (DVS) \cite{AS2019}; more specifically, a certain amount of reactive current from IBRs should be injected according to the local voltage at the point of coupling connection (PCC).

IBRs are considered promising for grid support due to the fast response and flexible controllability of power inverters, which motivates a number of research works on improving the low-voltage ride-through capability of IBRs. Today, the most prevalent strategy is the droop-based reactive current support \cite{YF2014,NT2015,MM2015}, which has also been specified as the DVS requirements in many grid codes. The droop control relies on the strong coupling between reactive power and voltage; therefore, it may not perform well for the grids with high R/X ratios and usually suffers from suboptimality.

To improve the DVS performance of inverters, several advanced control strategies have been proposed in recent years. In \cite{CA2014} and \cite{CA2021}, the (sequence or phase) voltage at the PCC is regulated to track a specific reference by injecting reactive current subject to the current limit of the inverter. The method in \cite{SMM2017} further maximizes the active power delivery while regulating the phase voltage within the pre-specified boundaries. In \cite{LXB021}, an optimal control method is developed to minimize the phase voltage deviation to the upper and lower limits while achieving specific active and reactive power delivery. For  \cite{CA2014,CA2021,SMM2017,LXB021}, a proper selection of voltage references and/or boundaries could be challenging and rather tricky in the sense that the depth of voltage sag is not \emph{apriori}. This may raise the risk of the infeasibility of control solutions and the optimality is not guaranteed.
\begin{figure*}[t]
    \centering
    \includegraphics[width=6in]{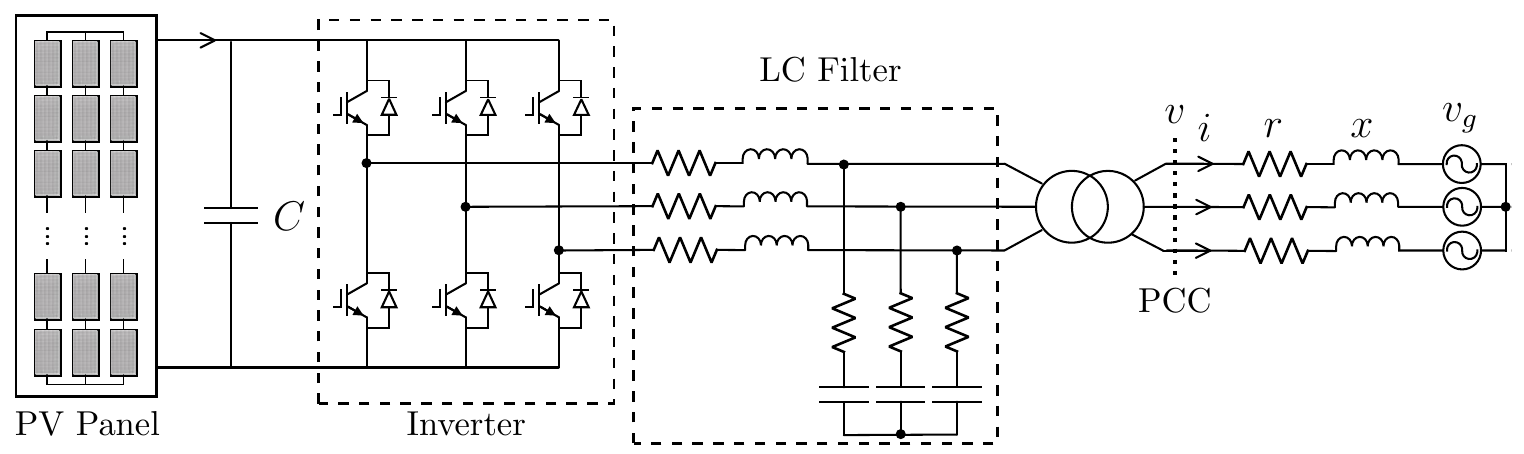}
    \caption{Schematic diagram of a three-phase grid-connected IBR system.}
    \label{schematicdiagram}
\end{figure*}
Another line of research directly optimizes the DVS performance while exploring the analytical solutions. In \cite{CA2016}, the authors propose an optimal DVS (ODVS) strategy to maximize the positive-sequence voltage while injecting the maximum fault current; three more flexible control objectives are considered in \cite{CA2018} and \cite{CA2018_2} to better deal with the unbalanced voltage sags and the active power oscillation reduction is further addressed in \cite{GM2020}. Since there is no evidence that the inverter should inject the maximum fault current, such equality constraint is relaxed to an inequality in \cite{SMA2019} to allow searching for the optimum on the whole feasible set rather than only on its boundary. A voltage-current sensitivity-based optimization model is proposed in \cite{KK2017}, wherein the maximum current injection is also mandatory. However, considering the difficulty of sensitivity computation in practice, the active and reactive currents are considered to be equal. Thus, a fixed control reference is implemented regardless of voltage sag conditions. 

Although the DVS control of IBRs has been well developed, several issues are still not clear so far and deserve further investigation:
\begin{itemize}
    \item Regarding
     droop control, there is still a lack of theoretical evidence for the (sub)optimality. 
    \item The active power limit of dc power source has not been directly considered in the existing ODVS schemes. Instead, several suboptimal strategies are developed to handle the power insufficiency issue \cite{CA2018,SMA2019}.
    \item Under deep voltage sag and/or weak grid conditions, the ``loss of synchronism" (instability) of inverters may occur \cite{GO2014,EI2009,WB2015,BJ2016}. However, this has  not been well addressed in the existing works. 
    \item The Karush–Kuhn–Tucker (KKT) conditions are used for solving the optimization programs---KKT point as optimum \cite{CA2016,CA2018,CA2018_2,GM2020,SMA2019}. However, KKT conditions are only valid for \emph{regular local optima}, and this has to establish on certain smoothness conditions \cite{BDP1999}. There is still no 
  evidence for the \emph{global optimality}, especially when the active power and stability limits are considered. 
\end{itemize}

Thus, the goal of this paper is to address all the above issues. We first revisit the droop control from an optimization perspective, and then develop a global optimal DVS (g-ODVS) controller based on the analytical optimality analysis. In detail, the contributions of this paper are outlined as follows. 
\begin{itemize}
    \item Firstly, the (sub)optimality of droop control is elaborated via the reverse engineering of optimization. This offers  insights into theoretical behaviors of  droop control.
    \item Secondly, a nonconvex optimization model to restore positive-sequence voltage is established, which considers the current limit and the active power and stability limits simultaneously.
    \item Thirdly, we, for the first time, present a rigorous optimality analysis that covers the existence, uniqueness, analytical expression, as well as  intuitive geometric illustration. All the KKT solutions, potential irregular solutions, and nondifferentiable solutions are checked to guarantee global optimality.
    \item Lastly, upon the optimality analysis, the g-ODVS controller is developed and implemented with a single-stage photovoltaic (PV) power plant. Its robustness analysis under model uncertainties is also provided.
\end{itemize}

The remainder of this paper is organized as follows. Section II presents the modeling of an inverter-based resource under voltage sags. Section III revisits the droop control from an optimization perspective. In Section IV, the optimal DVS control strategy is developed. Section V gives the implementation method with a PV system. Simulation results are presented in Section VI and robustness analysis is provided in Section VII, followed by the conclusion. Most proofs can be found in the arXiv version of this paper and are omitted.

\section{Problem Statement}
\iffalse
\begin{figure}
    \centering
    \includegraphics[width=2.3in]{equicirc.pdf}
    \caption{Equivalent circuit.}
    \label{equicirc}
\end{figure}\fi
\subsection{System Configuration}
Fig. \ref{schematicdiagram} shows the system configuration of a three-phase grid-connected inverter-based generator. The dc power source (PV, wind, energy storage, etc.) is connected into the grid through a voltage-source inverter, an inductance–capacitance (LC) filter and a step-up transformer, of which the grid side is considered as the PCC bus. 

\subsection{Inverter Modeling}
The typical control structure of a three-phase inverter consists of the outer voltage/power control loop and the inner current control loop under the synchronous dq-reference frame. To better deal with unbalanced operation conditions, the standard dual current control strategy that allows the decoupled control of positive-and negative-sequence currents is adopted. We refer the readers to \cite{YA2006} for more details about the dual current  control of inverters. 

The following assumptions are made  throughout the theoretical analysis.
1) This paper focuses on  positive-sequence voltage restoration. The negative-sequence active and reactive currents are considered zero, which can be accomplished by the dual  control. Hereinafter, any voltage, current and impedance refer to the physical quantities in the positive-sequence system, unless otherwise specified. 
2) Besides, the inner power losses of the inverter system are negligible. However, a real lossy system will be considered in the simulation.

Accordingly, the physics of inverters  can be described as:
\begin{subequations}\label{PhModel}
\begin{align}
    I&=\sqrt{I_{\rm d}^2+I_{\rm q}^2} \\
 V&=\sqrt{V_{\rm d}^2+V_{\rm q}^2}=V_{\rm d}\\ 
    P&=\dfrac{3}{2}\left(V_{\rm d}I_{\rm d}+V_{\rm q}I_{\rm q}\right)=\dfrac{3}{2}VI_{\rm d}\\
    Q&=\dfrac{3}{2}\left(V_{\rm q}I_{\rm d}-V_{\rm d}I_{\rm q}\right)=-\dfrac{3}{2}VI_{\rm q}
\end{align}
\end{subequations}
where $V, I, P$ and $Q$ are the voltage magnitude, current magnitude, active power and reactive power (in an average sense over a cycle) at the PCC; $V_{\rm d}$ and $V_{\rm q}$ are the d- and q-axis components of  PCC voltage; $I_{\rm d}$ and $I_{\rm q}$ are the active and reactive currents, respectively. The second ``$=$" in (\ref{PhModel}b)--(\ref{PhModel}d) holds when the d-axis is aligned to the positive-sequence voltage.

The current and active power limits of inverter are given as,
\begin{align}
 0\leq I&\leq \overline{I}\\
 \underline{P}\leq P&\leq \overline{P}
\end{align}
where $\overline{I}>0$ denotes the current limit; $ \underline{P}\leq0$ and $\overline{P}>0$ are the lower and upper active power limits that depend on dc-side energy resources (e.g., solar PV, wind, battery, etc). 
\subsection{Grid Modeling}
The  grid is modeled by a Thévenin equivalent  that emulates the voltage sag $(V_g,r,x)$ where $V_g$ is the grid voltage magnitude and $r+jx$ is the grid impedance. The grid parameters $V_g>0$, $r>0$ and $x>0$ are constant in the analysis. Note that the proposed methodology can also be  generalized to the purely resistive or purely inductive grids.

\subsection{Philosophy of Voltage Support }
 Based on Kirchhoff's voltage law, the  positive-sequence network can be expressed as:
\begin{align}
    \dot V-\dot V_g=\dot I(r+jx)\label{powerflowlaw}
\end{align}
where $\dot{V}_g:=V_ge^{j0}, \dot{V}:=Ve^{j\theta}$ and $\dot I:=Ie^{j\varphi}$ are the phasor representations of grid voltage, PCC voltage and output current, respectively.

Multiplying both sides of (\ref{powerflowlaw}) by $e^{-j\theta}$ and splitting the real and imaginary parts of the equation, it follows that
\begin{subequations}\label{powerflowreim}
\begin{align}
V-rI_{\rm d}+xI_{\rm q}&=V_g\cos{\theta}\\
rI_{\rm q}+xI_{\rm d}&=V_g\sin{\theta}.
\end{align}
\end{subequations}

Given that $-90^\circ\leq\theta\leq90^\circ$ for the sake of angle stability, it follows that
\begin{align}\label{Vexpression}
    V=\sqrt{V_g^2-(rI_{\rm q}+xI_{\rm d})^2}+rI_{\rm d}-xI_{\rm q}.
\end{align}

%There exists a \emph{unique} solution $(|V|,\theta)$, together with the given $(I_{\rm d},I_{\rm q})$, satisfying (\ref{powerflowreim}).

An underlying condition for (\ref{Vexpression}) is
\begin{align}
            |rI_{\rm q}+xI_{\rm d}|-V_g&\leq 0\label{stabilityconstr}
\end{align}
which guarantees a real value solution of $V$. This is equivalent to the stability conditions discussed in \cite{GO2014,EI2009,WB2015,BJ2016}. Several methods have been proposed to avoid this instability by modifying the droop control strategies \cite{GO2014,EI2009,WB2015,BJ2016}; however, its impact on ODVS has not been well addressed so far.

%(\ref{Vfeas}) is necessary because (\ref{powerflowreim}) is not strictly equivalent to the (\ref{powerflowlaw}), i.e., some solutions of (\ref{Vexpression}) is not feasible for (\ref{powerflowlaw}). More specifically, there exist solutions yielding $\left|V\right|<0$, which is obviously infeasible in a physical sense. 

%Obviously, $\left|V\right|$ and  $P$ can be considered as a function of $I_{\rm d}$ and $I_{\rm q}$, i.e., $\left|V\right|=v(I_{\rm d},I_{\rm q})$ and $P=p(I_{\rm d},I_{\rm q})$, respectively.

In summary, the inverter can be regarded as a  current source with independent control variables $I_{\rm d}$ and $I_{\rm q}$. DVS control aims to find the suitable $I_{\rm d}$ and $I_{\rm q}$  for improving $V$ subject to the current, active power, and stability limits.  

\section{Revisiting Droop Control}
Droop-based reactive current support strategies have been specified as the
DVS requirements in many grid codes. However, there is still a  lack of knowledge about its theoretical performance. So, in this section, we will revisit the droop control from an optimization perspective, which helps better understanding and evaluating its DVS behavior. 
\subsection{Preliminaries}
\begin{figure}
    \centering
    \includegraphics[width=3.5in]{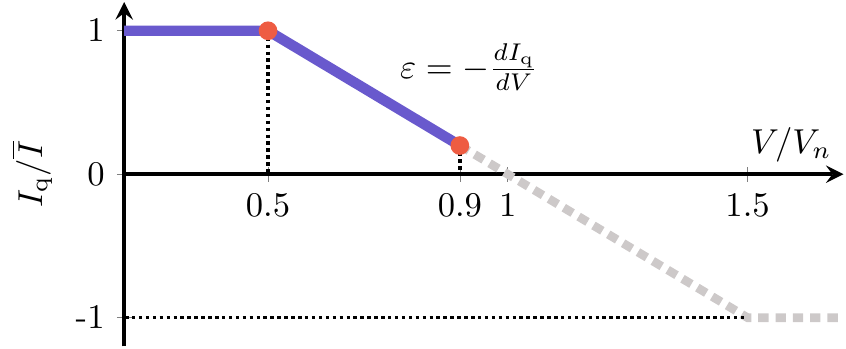}
    \caption{Droop rule for reactive current support in German grid codes \cite{AS2019}.}
    \label{drooprule}
\end{figure}
Without loss of generality, consider a typical piece-wise linear droop rule as in Fig. \ref{drooprule}, where the low-voltage section (solid line) is extended to construct a symmetrical rule with respect to $V=V_n$ for simplifying the mathematical expression. Note that this does not affect the results since the following analysis will only focus on the cases under sufficient voltage sags. Accordingly, the droop control rule can be described as, 
\begin{align}\label{droopcontrolrule}
    I_{\rm q}(t+1)=\Big[\varepsilon\cdot\big(V(t)-V_n\big)\Big]_{\mathcal{C}}
\end{align}
where $t$ is the time step; $V_n$ is the nominal voltage, typically 1.0 p.u.; $\varepsilon>0$ is the droop coefficient; $[\bullet]_{\mathcal{C}}$ denotes the projection onto the feasible set $\mathcal{C}:=\{I_{\rm q}|- \overline{I}\leq I_{\rm q}\leq\overline{I}\}$.

The following assumptions are made in this section.
\begin{description}[font=\normalfont]
     \item[A1] The active current is considered as zero, i.e., $I_{\rm d}=0$. Since most grid codes make no specific prescriptions regarding active current, the following analysis will be only concerned with the reactive current support.
     \item[A2] The voltage sag is not too severe and/or the grid is not too weak; more specifically, assume $\overline{I}<V_g/r$ holds. Without this assumption, there is a risk of  instability.
     \item[A3] The voltage cannot be fully restored to the nominal value $V_n$, i.e., ${\rm max}_{I_{\rm q}\in\mathcal{C}} V(I_q)<V_n$. In practice, this is widely believed to hold during grid faults.
\end{description}

The network behavior (\ref{Vexpression}) with the droop control rule (\ref{droopcontrolrule}) can be described as a nonlinear dynamical system,
\begin{align}\label{NDS}
\begin{array}{rl}
I_{\rm q}(t+1)=&\hspace{-3mm} \Big[\varepsilon\cdot\big(V(I_{\rm q}(t))-V_n\big)\Big]_{\mathcal{C}}.
\end{array}   
\end{align}

Thus, the equilibrium of the dynamical system $I_{\rm q}^\ast$ satisfies
\begin{align}\label{Equilibrium}
I_{\rm q}^\ast= \Big[\varepsilon\cdot\big(V(I_{\rm q}^\ast)-V_n\big)\Big]_{\mathcal{C}}.
\end{align}

    In what follows, the static performance of $I^\ast_{\rm q}$ will be evaluated from an optimization perspective.
\begin{figure*}
    \centering
    \includegraphics[width=2.5in]{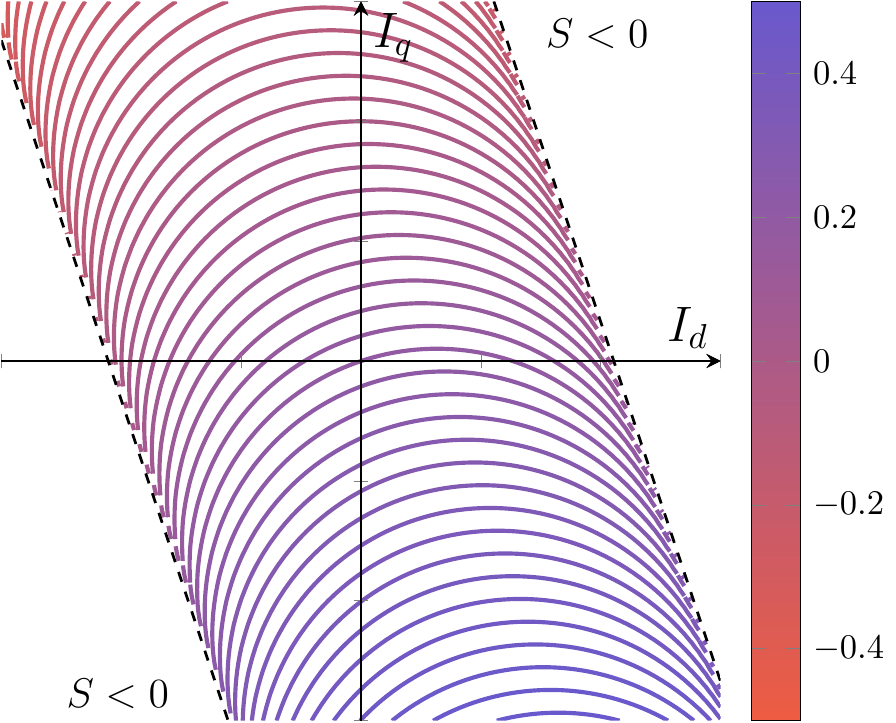}\hspace{30mm}
        \includegraphics[width=2.5in]{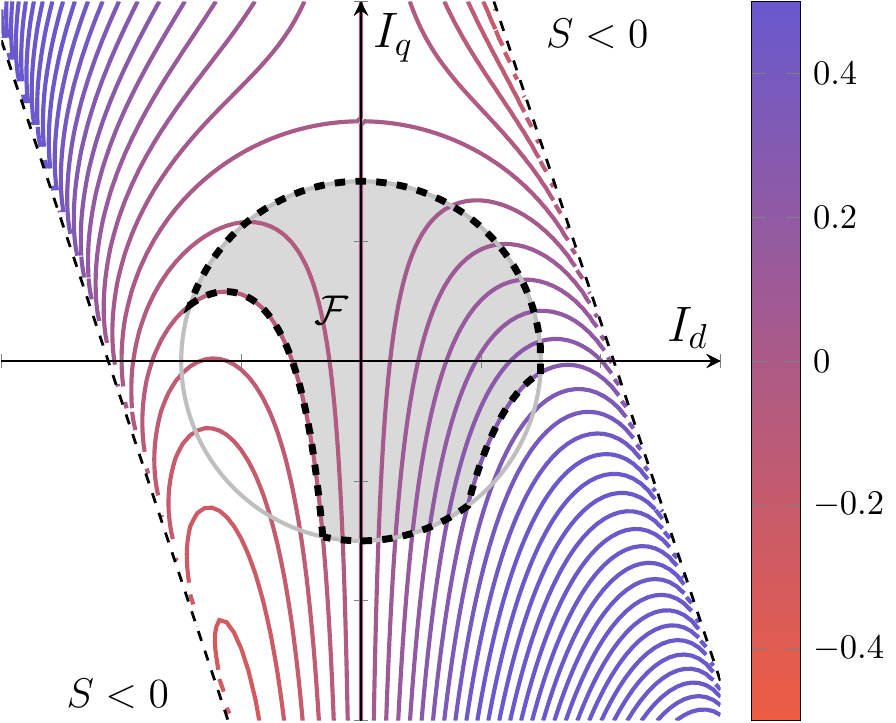}
    \caption{The geometry of the cost function and the constraint set of g-ODVS. The colored curves denote the contours of $V$ (in the left part) and $P$ (in the right part), the dashed black straight lines are the stability boundaries, i.e., $S=0$, the solid gray line (in the right part) is the current limit boundary, and the closed gray area is an example case of a nonconvex feasible set (in this example, $V_g=0.2 {\rm pu}$, $z=0.1 {\rm pu}$, and $r/x=1/3$).}
    \label{geometry}
\end{figure*}
\subsection{(Sub)optimality Analysis via Reverse Engineering}
It is difficult to directly evaluate the performance of an equilibrium point, although its explicit form can be obtained. So, we  take the approach of \emph{reverse engineering} \cite{CL2017}, of which the basic philosophy is: the equilibrium can be cast as a solution of a specific optimization program while the droop control can be interpreted as an online algorithm for solving the problem, i.e., network dynamics as optimization algorithms.

Given $I_{\rm q}^\ast$, define a function $F$:
\begin{align}\label{CostfcnRE}
   F(I_{{\rm q}}):=\frac{1}{2}\eta\big(V(I_{\rm q})-V_n\big)^2+\frac{1}{2\varepsilon}I_{{\rm q}}^2
\end{align}
 where $\eta$ is a constant defined as,
\begin{align}
    \eta:=-\frac{1}{\nabla V(I_{{\rm q}}^\ast)}.
\end{align}

So, $F$ is a function of $I_{\rm q}$ parameterized by the droop gain $\varepsilon$ and the constant $\eta$. Correspondingly, define the following optimization program,
\begin{align}\label{RE}
  {\bf RE:}\hspace{3mm} \underset{I_{\rm q}}{\rm min}\hspace{3mm} F(I_{\rm q})\hspace{3mm}
   \text{s.t.}\hspace{3mm}I_{\rm q}\in\mathcal{C}. 
\end{align}

Then, the DVS performance of the equilibria is elaborated in the following Theorems 1 and 2.

\emph{Theorem 1:} If $\overline{I}<xV_g/(rz)$, then the following statements are true:
\begin{enumerate}
    \item $I_{\rm q}^\ast$ is exactly the unique optimum of RE;
  \item $I_{\rm q}^\ast$  also maximizes $V$ if and only if $\varepsilon$ satisfies,
\begin{align}\label{SNC_droop}
    \varepsilon\geq\frac{-\overline{I}}{\sqrt{V_g^2-(r\overline{I})^2}+x\overline{I}-V_n}.
\end{align}
\end{enumerate}

In the case that $\overline{I}\geq xV_g/(rz)$,  $I_{\rm q}^\ast$ minimizes $F$, instead of minimizing $(V-V_n)^2$. Observe (\ref{CostfcnRE}), this achieves a compromise between the maximum voltage support and the minimum reactive current injection, which is characterized by the weights $\eta$ and $1/\varepsilon$. Upon A3, minimizing  $(V-V_n)^2$ equivalently  maximizes $V$ but the term $ I_{\rm q}^2/\varepsilon$ impedes the optimality.

\emph{Theorem 2:} If $\overline{I}\geq xV_g/(rz)$, the following statements are true:
\begin{enumerate}
    \item If $\nabla V(I_{\rm q}^\ast)<0$, $I_{\rm q}^\ast$ is the unique optimum of RE.  
    \item If $\nabla V(I_{\rm q}^\ast)>0$, $I_{\rm q}^\ast>-\overline{I}$, and $\nabla^2 F(I_{\rm q}^\ast)>0$, $I_{\rm q}^\ast$ is a local minimizer of RE.
    \item If $\nabla V(I_{\rm q}^\ast)>0$, $I_{\rm q}^\ast>-\overline{I}$ and $\nabla^2 F(I_{\rm q}^\ast)<0$, $I_{\rm q}^\ast$ is a local maximizer of RE.
    \item If $\nabla V(I_{\rm q}^\ast)=0$, $I_{\rm q}^\ast$ maximizes $V$. This  holds only if
    \begin{align}\label{NC_droop}
    \varepsilon=\frac{xV_g}{rzV_n-z^2V_g}.
\end{align}
\end{enumerate}

The case in Theorem 2 is much more complicated than that in Theorem 1 since $F$ is no longer guaranteed to be convex. Case 1) is the case as in Theorem 1.  Cases 2) and 3) achieve a local optimal trade-off between the  voltage restoration and  reactive current injection. In Case 4), $I_{\rm q}^\ast$ exactly achieves the optimal voltage support. 

The dynamical convergence process can be regarded as solving RE using a gradient-like algorithm with a varying step size.  The droop control usually achieves a compromise between the maximum voltage support and minimum reactive current injection. This sheds light on why droop control does not guarantee optimality, thereby motivating the advanced optimization-based strategies. 

\section{Optimal Dynamic Voltage Support}
We propose a g-ODVS control strategy, which aims to maximize the positive-sequence voltage while satisfying the current, active power and stability limits. It is formulated as,
\begin{subequations}\label{ODVS}
\begin{align}
\hspace{-5mm}{\bf g}\textbf{-}{\bf ODVS}:\hspace{3mm} \underset{I_{\rm d},I_{\rm q}}{\rm max}\hspace{3mm} &V(I_{\rm d},I_{\rm q})\\
{\rm s.t.}\hspace{3mm} & I(I_{\rm d},I_{\rm q})\leq\overline{I}\\
&P(I_{\rm d},I_{\rm q})\leq \overline{P}\\
&P(I_{\rm d},I_{\rm q})\geq \underline{P}\\
&S(I_{\rm d},I_{\rm q})\leq0\\
&V(I_{\rm d},I_{\rm q})\geq0
\end{align}
\end{subequations}
where (\ref{ODVS}b)--(\ref{ODVS}d) are the current, maximum power and minimum power limits, respectively; (\ref{ODVS}e) denotes the current angle stability limit [c.f. (\ref{stabilityconstr})]; (\ref{ODVS}f) is additionally required because (\ref{Vexpression}) does not guarantee $V\geq0$.  
Some basic properties of cost function and constraints are stated below:
\begin{itemize}
    \item $V$ is concave, $I$ is convex, $P$ is nonconvex and $S$ is linear.
    \item $V$ and $P$ are nondifferentiable at the stability boundary.
\end{itemize}

Therefore, g-ODVS is a nonlinear nonconvex program parameterized by $V_g, r, x, \overline{I}, \underline{P}$ and $\overline{P}$. The main complexity for solving it comes from the active power constraints.

\subsection{Geometry}
To better illustrate the characteristics of g-ODVS, its geometric illustration is given below. 

The left part of Fig. \ref{geometry} shows the contours of voltage $V$. From (\ref{Vexpression}), for given $U$, the trajectory of $V(I_{\rm d},I_{\rm q})=U$ is the half circle over the straight line $rI_{\rm d}-xI_{\rm q}-U=0$, with the center at $(Ur/z^2,-Ux/z^2)$ and  the radius equal to $V_g/z$. It is symmetric with respect to the straight line $rI_{\rm q}+xI_{\rm d}=0$.

The right part of Fig. \ref{geometry} gives the geometric illustration of constraints. The current limit $I(I_{\rm d},I_{\rm q})\leq\overline{I}$ is a closed circle area; $S(I_{\rm d},I_{\rm q})\leq0$ is the area between the two parallel lines. As observed in Fig. \ref{geometry}, the constraint set $\mathcal{F}$ becomes nonconvex only when $\underline{P}$ and/or $\overline{P}$ intersect with the current limit boundary. Note that if $V_g/z\leq\overline{I}$, the stability boundary $S(I_{\rm d},I_{\rm q})=0$ will intersect with the current limit boundary.

\subsection{Optimality Analysis }
The global optimum of g-ODVS is elaborated as follows.

\emph{Theorem 3:} g-OVDS has a unique optimum  $(I_{\rm d}^\star,I_{\rm q}^\star)$ with the following three stages:
\begin{enumerate}
    \item Stage 1 (S1): 
    \begin{align}\label{OPTS1}
         I_{\rm d}^\star=\dfrac{r}{z}\overline{I},\hspace{5mm}
         I_{\rm q}^\star=-\dfrac{x}{z}\overline{I}
    \end{align}
   if and only if the following condition C1 holds:
   \begin{align}
    \label{suffneceexact}
   {\rm C1:}\hspace{5mm}     \overline{P}\geq P_{b}:=\frac{3}{2}\left(\frac{r}{z}V_g\overline{I}+r\overline{I}^2\right).
   \end{align}
   \item Stage 2 (S2):
       \begin{align}\label{OPTS2}
       I_{\rm d}^\star=\overline{I}{\cos} \varphi^\star,\hspace{5mm}
       I_{\rm q}^\star=\overline{I}{\sin}\varphi^\star
    \end{align}
   where $\varphi^\star$ is the unique angle satisfying,
   \begin{align}\label{anglerangeS2}
       \left\{\hspace{-2mm}  \begin{array}{l}
       P\left(\varphi^\star\right)=\overline{P},\\[2mm]
       -90^\circ\leq\varphi^\star\leq{\rm atan2}\left(-\dfrac{x}{z},\dfrac{r}{z}\right)
             \end{array}\right.
   \end{align}
   if and only if neither of C1 and the following C2  hold:
   \begin{align}
  \hspace{-5mm} {\rm C2:}\hspace{2mm} \overline{I}\geq I_b:=&\sqrt{\frac{V_g^2}{2r^2}+\frac{2\overline{P}}{3r}+\frac{x^2-r^2}{2r^2z^2}V_g\sqrt{V_g^2+\frac{8r\overline{P}}{3}}}.
   \end{align}
   \item Stage 3 (S3): 
    \begin{align}\label{OPTS3}
         \left\{\hspace{-2mm}
         \begin{array}{l}
I_{\rm d}^\star=-\dfrac{1}{2z}\left(V_g-\sqrt{V_g^2+\dfrac{8r\overline{P}}{3}}\right)\\[5mm]
         I_{\rm q}^\star=-\dfrac{x}{2rz}\left(V_g+\sqrt{V_g^2+\dfrac{8r\overline{P}}{3}}\right)      
         \end{array}
         \right.
    \end{align}
    if and only if C2 holds.
\end{enumerate}

It is nontrivial to directly solve g-ODVS with all the constraints in an analytical way. However, based on some analysis resorting to the geometry, we can infer:
\begin{itemize}
    \item The optimum should be in the fourth quadrant ($I_{\rm d}>0,I_{\rm q}<0$). The current and maximum active power limits are critical, which are probably binding at the optimum. 
    \item If $\overline{P}$ is sufficiently large, the current limit $I(I_{\rm d},I_{\rm q})\leq\overline{I}$ will be the only critical constraint. 
    \item The stability constraint $S(I_{\rm d},I_{\rm q})\leq0$ is naturally hidden in the cost function and active power constraints. 
    \item The voltage limit (\ref{ODVS}f) will not be binding at the optimum since a feasible solution $V(0,0)=V_g>0$ always exists. 
\end{itemize}
 
Therefore, as will be detailed in the proof of Theorem 3, we resort to the exact relaxations of g-ODVS that are tractable instead of directly solving it.

Some interpretations of Theorem 3 are given as follows:
\begin{itemize}
    \item In S1, the inverter should inject the maximum current, accompanied by necessary active power curtailment. This solution is only related to $r/x$ and is not affected by the depth of voltage sag. The solution is equivalent to the optimal solution of \cite{CA2016,CA2018_2,SMA2019} but here we provide a rigorous condition for supporting it---this solution is valid if and only if C1 holds. Physically, S1 corresponds to the case that IBRs have high power productions, the voltage sag is deep and the maximum allowable current is low.  
    \item  In S2, the maximum active power and the maximum current are achieved, simultaneously. It is only valid if neither of C1 and C2 hold---a relatively low power production and/or a moderate voltage sag.
  Interestingly, the solution in S2 is essentially close to the suboptimal strategies in \cite{CA2016,SMA2019} but  different in terms of implementation method, which will be detailed later. Besides, no explicit expression is available for the solution in  S2.
    \item In S3, the inverter should inject the maximum available  power but the injected current magnitude should not attain its limit.  S3 is usually induced by deep voltage sags, weak grids, high overcurrent capabilities of inverters, and low power productions of IBRs. Note that, unlike S1 and S2, S3 will never happen under certain conditions regardless of the amount of available power. To be clear, a necessary condition for S3 is:
    \begin{align}\label{necessaryS3}
 {\rm C3:}      \hspace{5mm} \frac{x}{r}\cdot\frac{V_g}{z}<\overline{I}
    \end{align}
  which is only related to the grid voltage, short-circuit ratio (SCR), $r/x$ ratio and current limit. Upon C3, a variant of C2 regarding active power limit is given as:
   \begin{align}\label{C3}
     \hspace{-5mm}  \nonumber{\rm C2}^{\prime}: \overline{P}< P_b^\prime:=\frac{3}{8r}\Bigg(&\Bigg(\frac{r^2-x^2}{z^2}V_g\\
       &+2r\sqrt{\overline{I}^2-\frac{x^2V_g^2}{z^4}}\Bigg)^2-V_g^2\Bigg).
   \end{align}
   % More details are referred to the proof of Theorem 3 in Appendix.
\end{itemize}
\iffalse
\begin{table}[t]
\small
\centering
\caption{System Parameters}\label{mainparameters}
\renewcommand\arraystretch{1.2}
\begin{tabular}{rccc}
        \hline\hline
Factor &S1&S2&S3\\
         \hline
Grid voltage $V_g$&$-$&$-$&$-$ \\
$r/x$ ratio&$+$&&\\
SCR ($1/z$)&$-$&&\\
Power limit $\overline{P}$&$+$&&\\
Current limit $\overline{I}$&$-$&&\\
 \hline\hline
        \end{tabular}
\end{table}
\fi

\subsection{Geometric Illustration}
\begin{figure*}
    \centering
    \includegraphics[width=2.1in]{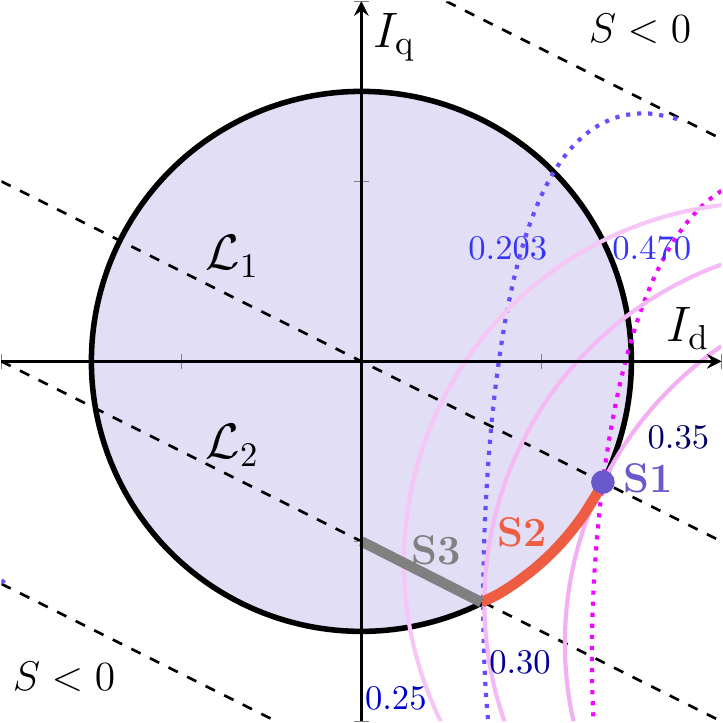}\hspace{40mm}
     \includegraphics[width=2.1in]{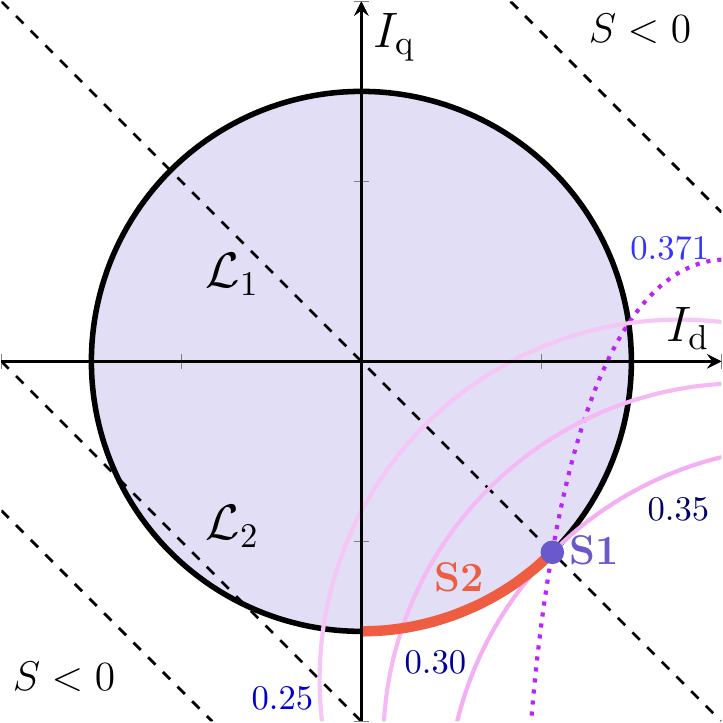}
    \caption{Trajectories of optimal solution where the solid contours correspond to the PCC voltage $V$ and the dotted contours correspond to the  active power $P$. Left: $V_g=0.2{\rm pu}, z=0.1{\rm pu}$ and $r/x=2$; Right: $V_g=0.2{\rm pu}, z=0.1{\rm pu}$ and $r/x=1$.}
    \label{traj}
\end{figure*}
To better interpret Theorem 3, we give some geometric illustration on the trajectory of optimal solutions under different   settings of $\overline{P}$. 
Fig. \ref{traj} shows two example cases. In the first case (left-hand side in Fig. \ref{traj}), the trajectory of optima is S1(violet  point)$\rightarrow$S2(thick orange line)$\rightarrow$S3(thick gray line). If $\overline{P}>P_b$, the optimum will be  the unique point S1. It is the intersection between the current limit boundary and the straight line 
\begin{align}
    \mathcal{L}_1: \hspace{3mm}rI_{\rm q}+xI_{\rm d}=0
\end{align}
 in the fourth quadrant. If $\overline{P}$ gradually decreases, such that $\overline{P}<P_b^\prime$, the optimum moves along the current limit boundary (in the direction that $I_{\rm d}$  decreases monotonically) and finally reaches the intersection point  between the current limit boundary and the straight line: 
\begin{align}\label{dVdIq}
  \mathcal{L}_2:\hspace{3mm}\frac{\partial V}{\partial I_{\rm q}}=0\Longrightarrow rI_{\rm q}+xI_{\rm d}+\frac{x}{z}V_g=0.
\end{align}
If $\overline{P}$ further decreases, the optimum will move inside the current limit circle along $\mathcal{L}_2$. Correspondingly, the achieved voltage from S1 to S3 decreases as $\overline{P}$ decreases. 

In the second case (right-hand side in Fig. \ref{traj}), the trajectory of optima is S1$\rightarrow$S2. Here, S3 does not exist anymore since there is no intersection between $\mathcal{L}_2$ and the current limit boundary in the fourth quadrant, i.e., C3 does not hold.

\section{Implementation}
This section is devoted to the implementation method of g-ODVS with a grid-connected single-stage PV power plant. The g-ODVS controller will be integrated into the IBR control system to generate current references for voltage support.
\subsection{Overview}
In normal operation, the PV system operates in the maximum power point tracking (MPPT) mode with unity power factor. The perturb \& observe technique is used for MPPT \cite{PandOMPPT}.  When the voltage sag is detected (e.g., $V\leq0.85$ pu), it switches to the voltage support mode with the following tasks:
\begin{itemize}
    \item The dc voltage reference from the MPPT controller and the available power measured at the PCC are latched while the normal dc voltage control is locked. Note that, we take the pre-fault power measurement at the PCC as the available power, which is based on the assumption that the system operates with the MPPT mode before the voltage sag occurs. Such feedback can satisfactorily account for the posterior power losses inside the PV system. 
    \item The grid model estimation is triggered, which provides the grid parameters for  the g-ODVS controller. The following simple strategy is used to estimate the grid voltage. During the first $m$ cycles ($m=3\thicksim5$) after triggering g-ODVS, the positive and negative sequence $\rm dq$ current references are set as zero. Thus, the grid voltage is equal to the PCC voltage, so that it can be locally estimated. As for the grid impedance, it is assumed to be available in this work. However, in practical implementation, the online grid estimation is preferred.
    \item Based on the knowledge of grid conditions ($V_g,r,x$) and PV system parameters ($\overline{I},\overline{P}$), the g-ODVS controller generates current references  according to Theorem 3.
\end{itemize}
\begin{figure}[t]
    \centering
    \includegraphics[width=2.45in]{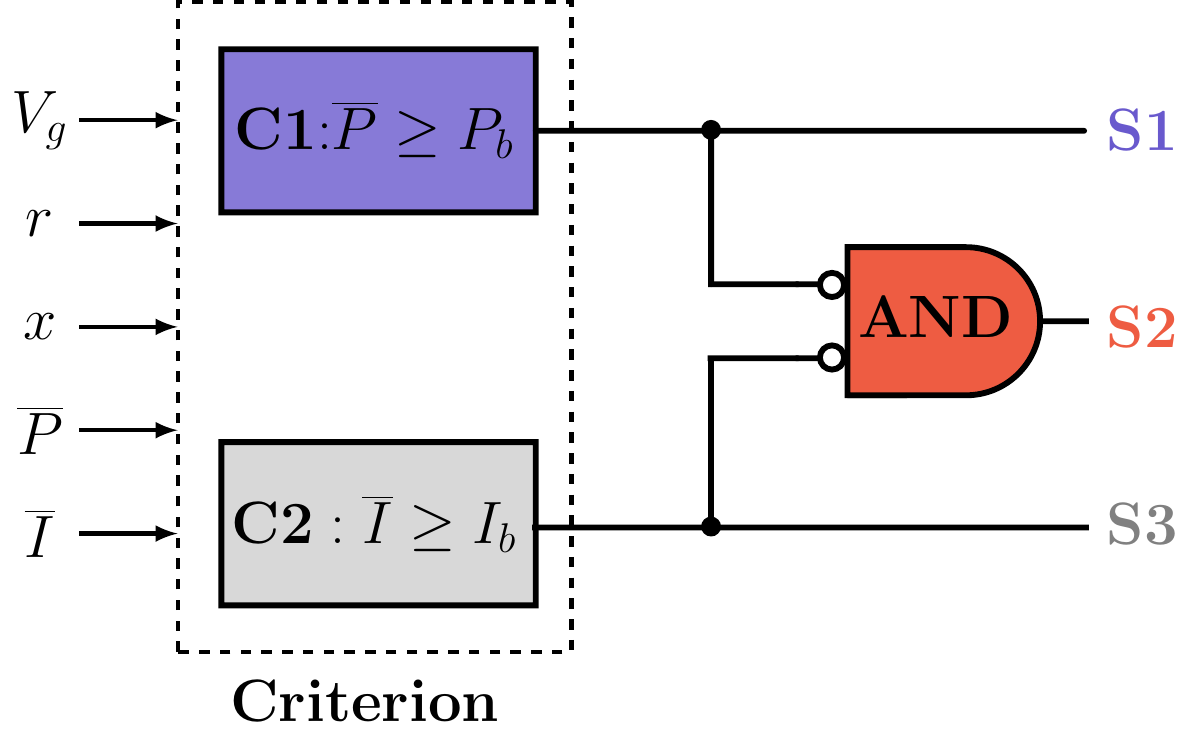}
    \caption{Logic loop for stage decision.}
    \label{logic}
\end{figure}
\subsection{G-ODVS Controller Design}
\begin{figure*}
    \centering
    \includegraphics[width=5.8in]{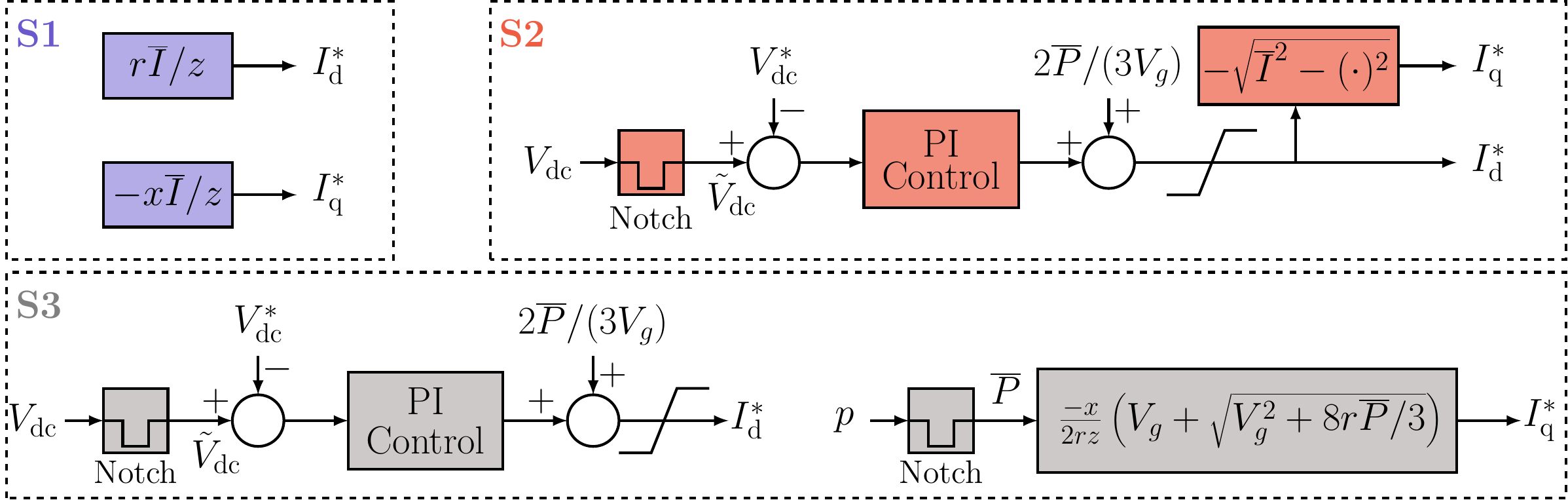}
    \caption{G-ODVS controller design.}
    \label{ODVScontroller}
\end{figure*}
The logic loop based on Theorem 3 that determines which stage  is active is illustrated in Fig. \ref{logic}. The g-ODVS controller design is shown in Fig. \ref{ODVScontroller}, which is described as follows: 
\begin{itemize}
    \item In S1, since the explicit closed-form solution (\ref{OPTS1}) is available,  we can directly set the $dq$ current references.
    \item In S2, no explicit expression of the optimum is available but according to (\ref{OPTS2})--(\ref{anglerangeS2}), the optimum should be on the current limit boundary within a specific range and  the maximum active power is delivered. Thus, a simple strategy is designed to search for the solution in a closed-loop way: $I_{\rm d}$ is regulated to track the maximum active power using a PI-based dc voltage controller with an anti-windup strategy and $I_{\rm q}=-\sqrt{\overline{I}^2-I_{\rm d}^2}$. As long as the maximum active power output is attained, the optimality is achieved. In spirit, S2 is close to the suboptimal strategies in \cite{CA2018_2,SMA2019}; however, the suboptimal strategies generate the reference of $I_{\rm q}$ by a reactive power control loop. %The suboptimal strategies in \cite{CA2018_2, SMA2019} estimates the available reactive power based on the voltage measurement feedback and current limit. In this way, it is essentially equivalent to our proposed controller, though it is developed under $\alpha\beta$ reference frame. 
    In other words, this paper proves that the so-called ``suboptimal" strategies in \cite{CA2018_2,SMA2019} can achieve the optimality if C2 (or equivalent C2$^\prime$) holds. %However, the suboptimal strategy in \cite{SMA2019} computes the available reactive power based on the rated capacity, which may lead to over-current during voltage sags if without current limiting.
    \item As for S3, although the explicit solution (\ref{OPTS3}) is available, for better robustness against model errors, we combine the closed-loop and open-loop control. For $I_{\rm d}$, we adopt the same strategy as in S2 since the maximum active power constraint is also binding in this stage. The  $I_{\rm q}$ reference is directly computed according to (\ref{OPTS3}b) using the feedback measurement of $\overline{P}$ [or equivalently computed by (\ref{dVdIq}) using the feedback of $I_{\rm d}$]. In this way, the stability of dc-link voltage can be guaranteed with model errors. In contrast, a full open-loop control may lead to dc voltage collapse because the open-loop control solution (\ref{OPTS3}) under erroneous grid models may result in larger active power outputs than the available power.
%    \item As mentioned above, the negative-sequence (dq) current references are set as zero.
\end{itemize}
 
\section{Simulation Results}

\iffalse
\begin{figure}
    \centering
    \includegraphics[width=3in]{PVcurve.pdf}
    \caption{P-V characteristics of PV array.}
    \label{PVcurve}
\end{figure}
\fi
In this section, we test the proposed g-ODVS with a grid-connected PV system (as shown in Fig. \ref{schematicdiagram}) via dynamic simulations in MATLAB/Simulink R2021a environment. The main specifications of the test system are given in Table \ref{mainparameters}. 
The PV array consists of 88 strings of  modules (SunPower SPR-415E-WHT-D \cite{Sunpower}) connected in parallel; each string consists of 7 modules connected in series. %of which the P-V characteristics is shown depicted in Fig. \ref{PVcurve}. 
%More details of PV module are referred to  \cite{Sunpower}.
Three cases are considered for testing our proposal under different operation conditions. We compare g-ODVS with the conventional droop control (with reactive current priority) and the (sub)optimal DVS control strategy (ODVS for short) in \cite{CA2018_2,SMA2019}. 
\iffalse The droop control rule used for comparison is,
\begin{align}\
\nonumber I_{\rm q}=\left\{\hspace{-2mm}
   \begin{array}{ll}
    -\overline{I},& {\rm if}\,V\leq0.5\\[0mm]
    \dfrac{\overline{I}}{0.4}(V-0.9),    & {\rm if}\,0.5<V<0.9\\[2mm]
    0,& {\rm if}\,V\geq0.9 
   \end{array},
   \right.
   I_{\rm d}=\sqrt{\overline{I}^2-I_{\rm q}^2}.
\end{align}
\fi
\begin{table}[t]
\small
\centering
\caption{System Parameters}\label{mainparameters}
\renewcommand\arraystretch{0.95}
\begin{tabular}{ m{2in} m{1in}}
        \hline\hline
Description&Value\\
         \hline
Power rating& 250 kW (1 pu) \\
Nominal ac voltage& 250V/25kV (1 pu)\\
Nominal dc voltage&480V\\
Nominal frequency& 60 Hz\\
Pre- and post-fault SCRs& 20/10\\
R/X ratio&2\\
Maximum current limit& 1.5 pu\\
Number of delayed cycles $m$& 3\\
%PV module& SunPower SPR-415E-WHT-D \\
 \hline\hline
        \end{tabular}
\end{table}

\subsection{Moderate Voltage Sag With High Power Production}
\begin{figure}
    \centering
        \includegraphics[width=3.5in]{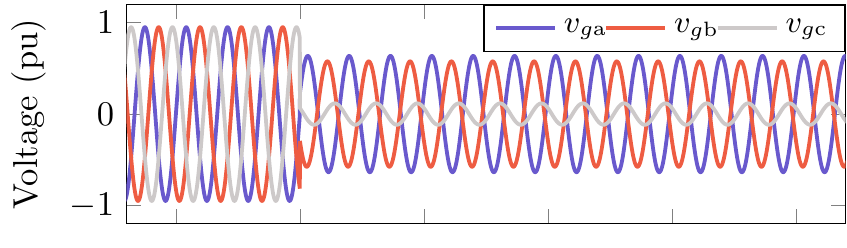}
    \includegraphics[width=3.5in]{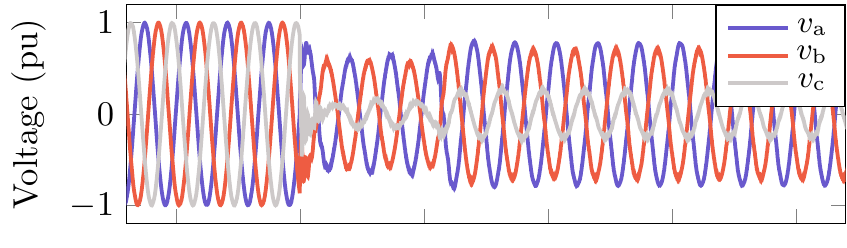}
    \includegraphics[width=3.5in]{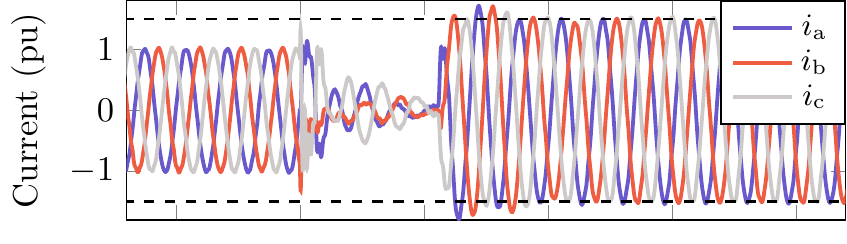}
      \includegraphics[width=3.5in]{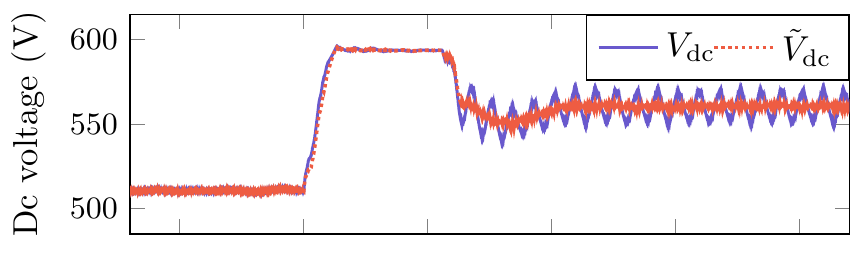} \includegraphics[width=3.5in]{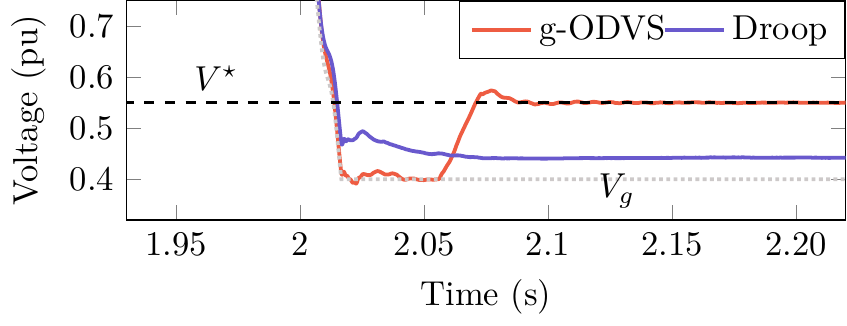}
    \caption{Simulation results under Case A (from top to bottom: grid voltage, PCC voltage, output current, dc voltage, and PCC positive-sequence   voltage).}
    \label{figs_caseA}
\end{figure}

In Case A, unbalanced voltage sag is imitated where the positive-sequence voltage is $0.4\angle0^\circ$pu  and the negative-sequence grid voltage is set as $0.3\angle50^\circ$pu. The solar irradiance is $1000\rm W/m^2$ and the cell temperature is $25^\circ\rm C$. Since $\overline{P}=241.4 {\rm kW}>184.4{\rm kW}=P_b$, the inverter should operate with S1. The simulation results are depicted in Fig. \ref{figs_caseA}. The voltage sag occurs at $t=2$s and the g-ODVS controller is triggered at $t=2.0065$s. After the three-cycle ($0.05$s) delay for grid parameter detection (the estimated $V_g$ is $0.3998$pu at the end of the third cycle), the optimal solution is implemented and then $V$ is recovered to $0.55$pu after several tens of milliseconds. The resultant positive-sequence voltage is exactly equal to the theoretical optimum. 
In comparison, the droop control achieves a less good control performance in this case where the voltage is recovered to $0.44$pu. Keep in mind that ODVS generates the same control solution as g-ODVS in S1 and S2, which is thus omitted here. As for the dc-side dynamics, the single-stage PV system is self-protected because the power generation of the PV array is reduced when the dc voltage increases under ac voltage sags. Therefore, although the resultant active power output is lower than the available power, such inherent self-protection capability can drive the PV generation system towards a new equilibrium (the filtered dc voltage $\tilde{V}_{\rm dc}=560$V). However, for a two-stage PV system that is not self-protected, the necessary
supplemental control is required to coordinate the dc/dc converter with the g-ODVS controller so that the PV system can maintain stability.
\subsection{Moderate Voltage Sag With Medium Power Production}
\begin{figure}
    \centering
       \includegraphics[width=3.5in]{vgabc_caseAB.pdf}
    \includegraphics[width=3.5in]{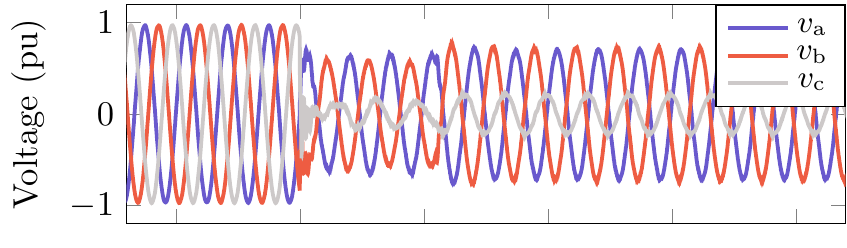}
    \includegraphics[width=3.5in]{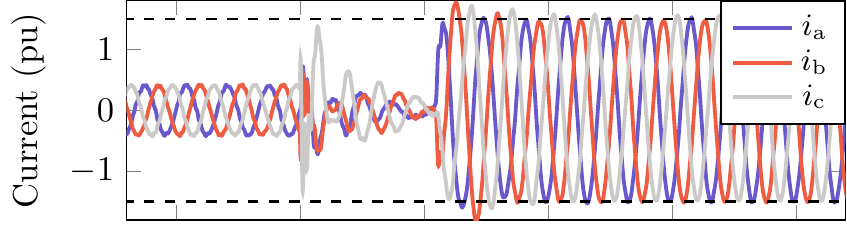}
      \includegraphics[width=3.5in]{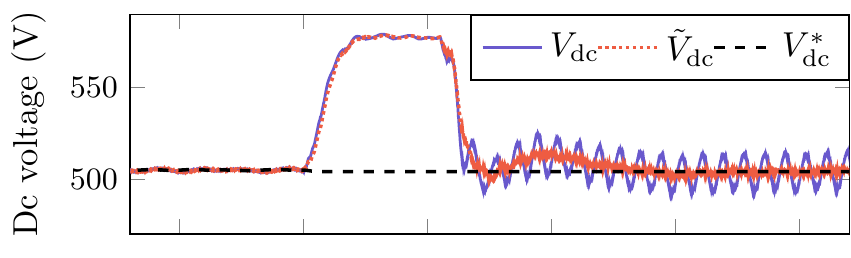} \includegraphics[width=3.5in]{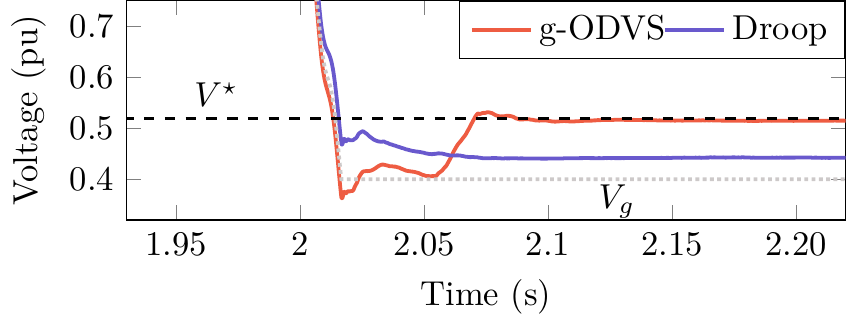}
    \caption{Simulation results under Case B (from top to bottom: grid voltage, PCC voltage, output current, dc voltage, and PCC positive-sequence  voltage).}
    \label{figs_caseB}
\end{figure}
In Case B, the grid condition is the same as that in Case A; the solar irradiance is $400\text{W}/\text{m}^2$ and the cell temperature is  $25^\circ\rm C$. $\overline{P}=95.4\text{kW}<185.4\text{kW}=P_b$ and $I_b=2.48\text{pu}>\overline{I}$. Therefore, the PV inverter should operate with S2. The simulation results are depicted in Fig. \ref{figs_caseB}. With the g-ODVS control strategy, the positive-sequence voltage is restored to $0.5145$pu, which is very close to the theoretical optimum $0.5157$pu obtained by solving the nonlinear equation group $I(I_{\rm d},I_{\rm q})=\overline{I}$ and $P(I_{\rm d},I_{\rm q})=\overline{P}$. The slight deterioration is caused by the power losses inside the PV generation system and the estimation error of grid voltage ($0.4026$pu).
Owing to the limited available power (i.e., C1 does not hold), the resultant voltage is lower than that in Case A. The dc voltage tracks the reference immediately after the optimal solution is implemented. %As discussed before, the suboptimal solution in ODVS \cite{CA2018_2,SMA2019} is equivalent to S2.

\subsection{Deep Voltage Sag With Low Power Production}
\begin{figure}
    \centering
    \includegraphics[width=3.5in]{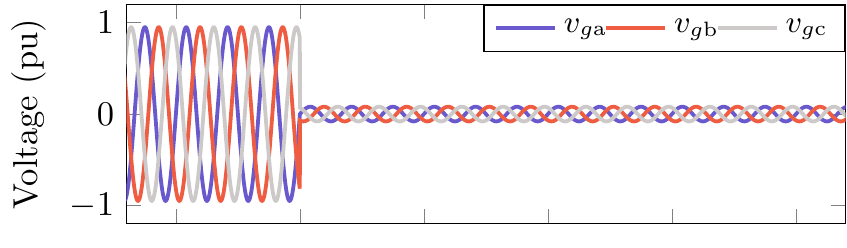}
    \includegraphics[width=3.5in]{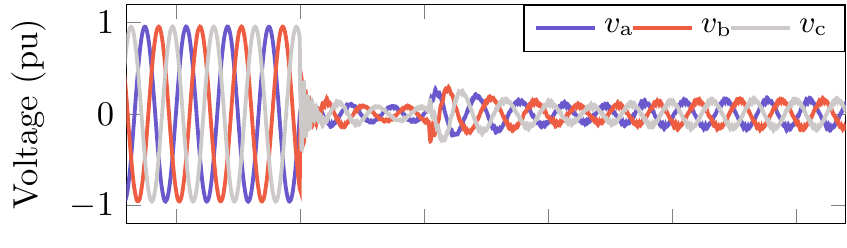}
    \includegraphics[width=3.5in]{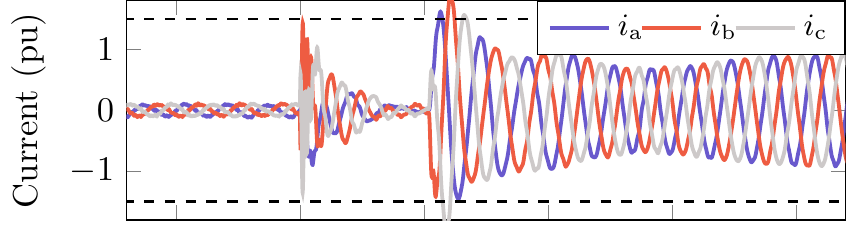}
      \includegraphics[width=3.5in]{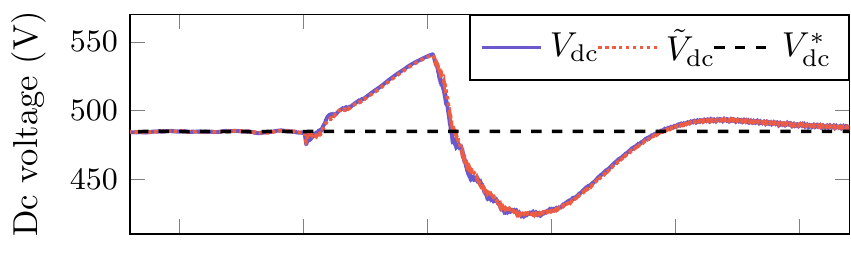} \includegraphics[width=3.5in]{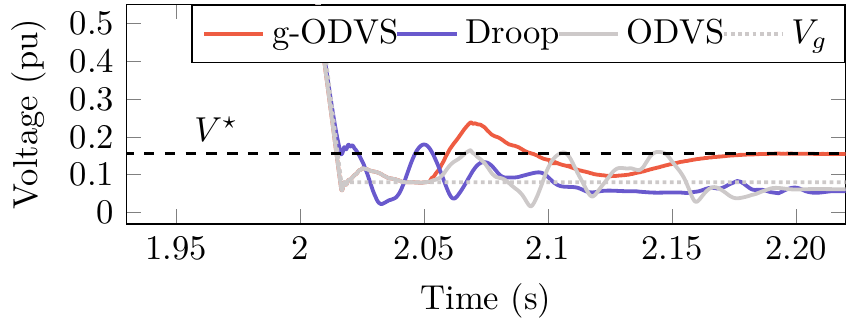}
    \caption{Simulation results under Case C (from top to bottom: grid voltage, PCC voltage, output current, dc voltage, and PCC positive-sequence   voltage).}
    \label{figs_caseC}
\end{figure}
\begin{figure}
    \centering
    \includegraphics[width=3.45in]{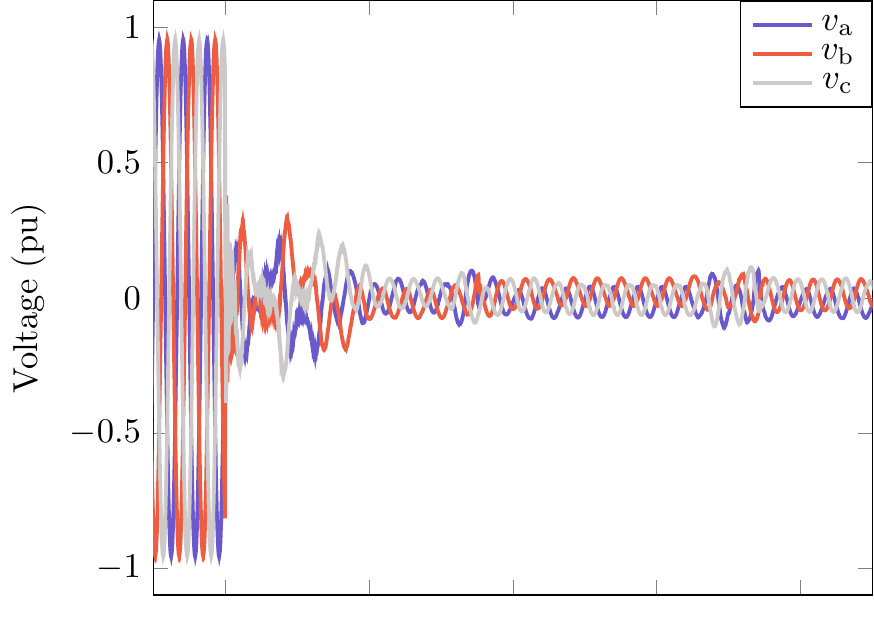}
    \includegraphics[width=3.45in]{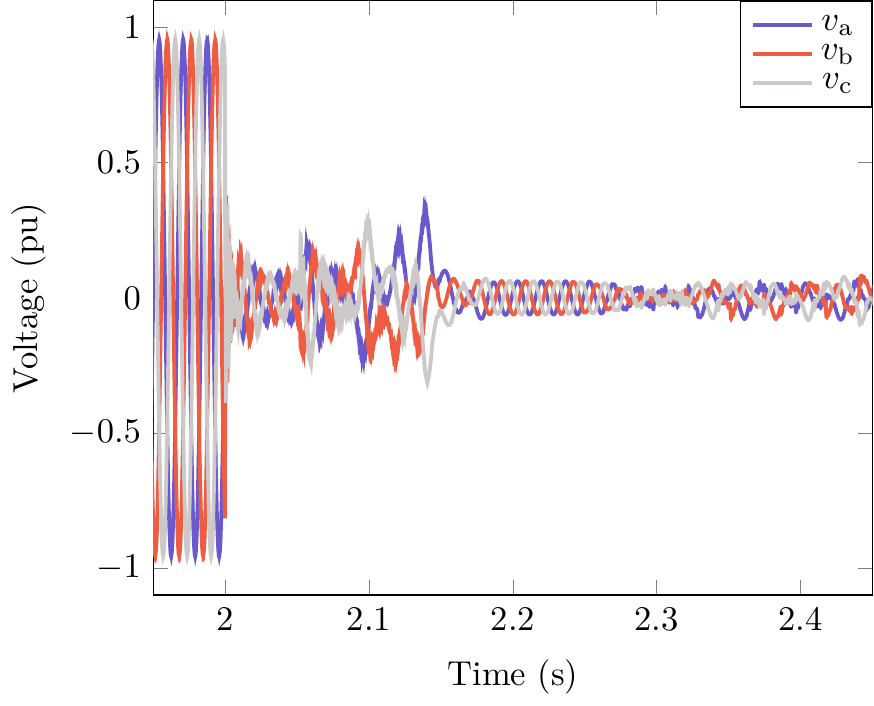}
    \caption{PCC voltage under droop control (top) and ODVS (bottom).}
    \label{distoredvoltage}
\end{figure}
In Case C, a severe balanced voltage sag is simulated
where $V_g=0.08\angle0^\circ$pu. The solar irradiance is $100\text{W}/\text{m}^2$ and the cell temperature is set as $25^\circ\rm C$. The estimated grid voltage is $0.0806$pu. Given that $\overline{P}=23.1\text{kW}<77.4\text{kW}=P_b$ and $I_b=0.9145\text{pu}<\overline{I}$, the inverter operates with S3 for g-ODVS. The droop control and ODVS (operates with the suboptimal mode) result in significantly distorted voltage (see Fig. \ref{distoredvoltage}) because of the loss of synchronism. In other words, droop control and ODVS drive the system towards an operating point that is physically infeasible. However, in comparison, g-ODVS can maintain stability  while recovering the voltage to approximately $0.1532$pu, which is slightly lower than the estimated optimum $0.1557$pu. The dc voltage  is restored to the reference value after $t=2.15$s.

\section{Robustness Analysis}
In the dynamic simulations, the accuracy of  grid voltage and active power estimation has been well demonstrated. However, the grid impedance estimation is assumed to be perfect, which may not be the case in practice. Therefore, in this section, we will address the impact of erroneous grid impedance. 

The mismatch of grid impedance could lead to mistakes in the stage decision and  suboptimality even though the stage is correctly selected. The former situation rarely happens because it requires a sufficiently large error while the available active power is very close to the boundary in C1 or C2$^\prime$. Thus, we only focus on its impact on the \emph{optimality gap} here.

The uncertainty of impedance estimation is modeled by,
\begin{align}
    \hat{r}&=(1+\alpha)r,\hspace{3mm}\alpha\in[\underline{\alpha},\overline{\alpha}]\\
    \hat{x}&=(1+\beta)x,\hspace{3mm}\beta\in[\underline{\beta},\overline{\beta}]
\end{align}
where $[\underline{\alpha},\overline{\alpha}]$ and $[\underline{\beta},\overline{\beta}]$ are the independent error bands for grid resistance and reactance, respectively.

{\bf S1).}
The solution in S1 is implemented in an open-loop manner as in (\ref{OPTS1}). It varies with $\hat r/\hat x$ but
 should  always be on the current limit boundary around the true optimum. Define $\Delta\varphi:={\rm atan2}(-\hat{x}/\hat{z}, \hat{r}/\hat{z})-{\rm atan2}(-{x}/{z},{r}/{z})$, it follows that the optimality gap $G:=V^\star-\hat V^\star$ monotonically increases as $|\Delta\varphi|$ increases. Therefore, the theoretical upper bound of the optimality gap $\overline{G}$ under the uncertainty is given by,
\begin{align}
\overline{G}=V^\star-{\rm min}\left\{\hat V^\star(\underline{\alpha},\overline{\beta}),\hat V^\star(\overline{\alpha},\underline{\beta})\right\}
\end{align}
where
\begin{align*}
    \hat V^\star(\underline{\alpha},\overline{\beta})&=V\left(\frac{\overline{I}}{\sqrt{1+\frac{(1+\overline{\beta})^2}{(1+\underline{\alpha})^2}\frac{x^2}{r^2}}},\frac{-\overline{I}}{\sqrt{1+\frac{(1+\underline{\alpha})^2}{(1+\overline{\beta})^2}\frac{r^2}{x^2}}}\right)\\
     \hat V^\star(\overline{\alpha},\underline{\beta})&=V\left(\frac{\overline{I}}{\sqrt{1+\frac{(1+\underline{\beta})^2}{(1+\overline{\alpha})^2}\frac{x^2}{r^2}}},\frac{-\overline{I}}{\sqrt{1+\frac{(1+\overline{\alpha})^2}{(1+\underline{\beta})^2}\frac{r^2}{x^2}}}\right).
\end{align*}

\begin{figure}[t!]
    \centering
    \includegraphics[width=3.5in]{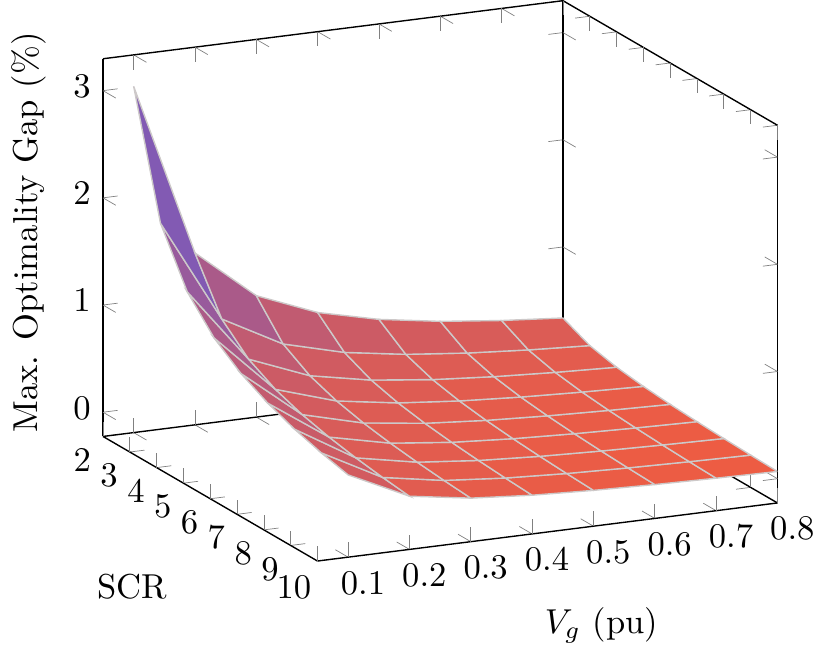}
    \caption{Maximum optimality gap of S1.}
    \label{OptGapS1}
\end{figure}

Fig. \ref{OptGapS1} gives the numerical results of the relative maximum gap ($\%\overline{G}=\overline{G}/V^\star\times100\%$) under different $V_g$ and SCR conditions, where   $\underline{\alpha}=\underline{\beta}=-0.1$, $\overline{\alpha}=\overline{\beta}=0.1$, $r/x=2$ and $\overline{I}=1.5$pu. It can be observed that $\%\overline{G}$ is less than $3\%$ in a wide regime. This indicates that g-ODVS is robust in S1.

{\bf S2).} As for S2, 
the solution is implemented in a fully closed-loop manner. Therefore, the optimality is not affected by the grid impedance estimation, i.e., $G=0$.

\begin{figure}[t!]
    \centering
    \includegraphics[width=3.5in]{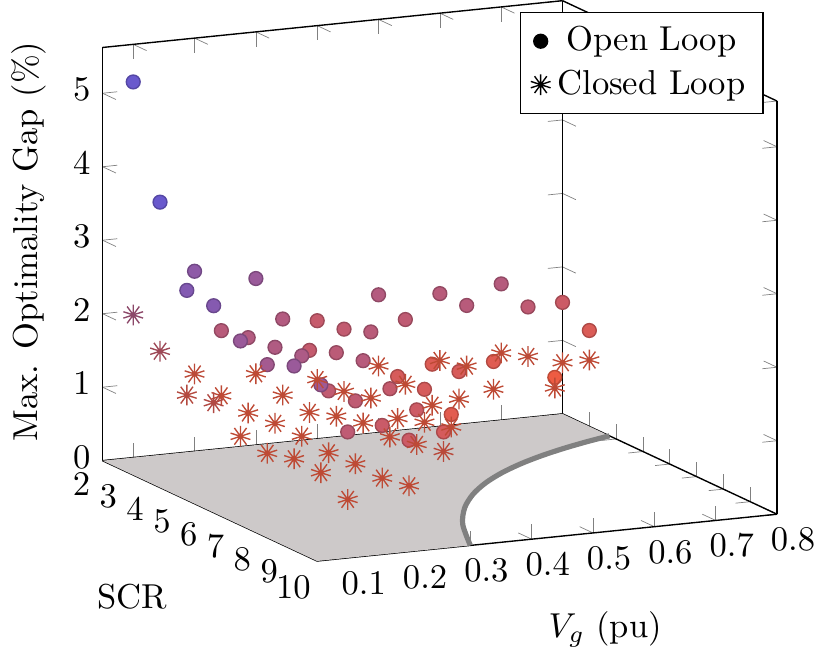}
    \caption{Maximum optimality gap of S3.}
    \label{OptGapS3}
\end{figure}

{\bf S3).} The solution in S3 can be implemented either in a fully open-loop  manner [by (\ref{OPTS3})] or a (half) closed-loop manner (as in Fig. \ref{ODVScontroller}). It is hard to analytically find the maximum optimality gap for S3. 
So, we leverage the Monte-Carlo method to estimate the gap bounds under different grid voltage ($0.1,0,2,...,0.8$pu) and SCR ($2,3,...,10$) conditions. For each case  (a combination of $V_g$ and SCR) satisfying C3, i.e., within the gray area in Fig. \ref{OptGapS3}, 200 Monte-Carlo trails are carried out where $\alpha$ and $\beta$ are sampled from the uncertainty set following the uniform distribution. According to C2$^\prime$, each case has a threshold $P_b^\prime$ corresponding to the transition point between S2 and S3. To reduce the risk of overcurrent under uncertainty, we take $\overline{P}=P_b^\prime/2$ in the tests. Fig. \ref{OptGapS3} shows the estimated $\%\overline{G}$, where in the open-loop control test, the random samples that result in $P>\overline{P}$ are not taken into account since they are physically infeasible due to power imbalance. The results suggest that S3 also exhibits good robustness
against grid impedance estimation errors. The closed-loop control performs better than the open-loop counterpart ($\%\overline{G}=1.95\%$ v.s. $\%\overline{G}=5.12\%$).


\section{Conclusion}
This paper investigates the voltage support control of inverters under voltage sags. We first theoretically elaborate why the droop control usually only provides suboptimal performance. Then, we propose a global optimal control strategy.
A nonconvex optimization model g-ODVS is established, which maximizes the positive-sequence voltage recovery, considering the current, active power and stability limits. The program is then  solved  analytically and rigorously, so that global optimality is guaranteed. It is proven that the unique optimum has three different stages depending on specific conditions---C1 and C2. The optimum reveals that
\begin{enumerate}
    \item active power should be given higher priority than reactive power especially when the available active power is not sufficient (S2 and S3); 
\item under severe voltage sags and very low available power (S3), inverters should \emph{not} inject the maximum allowable current; otherwise, in addition to suffering from suboptimality, there is a risk of loss of synchronism (instability). 
\end{enumerate}

The optimum is implemented with a grid-connected single-stage PV power plant. The simulation results demonstrate that g-ODVS can better guarantee both optimality and stability, compared with the conventional droop control and existing (sub)optimal control strategies. Moreover, the robustness of optimality against model mismatch is also examined. The numerical experiments verify that the  optimality gap is relatively small ($<3\%$) under $10\%$ impedance estimation error, suggesting that  the proposed g-ODVS controller is robust.

\appendices
\section{Proofs of Theorem 1 and Theorem 2}
\subsection{Preliminaries}
For brevity, let us define the critical condition that results in two different cases in Theorems 1 and 2, respectively,
\begin{align}
{\rm C4:} \hspace{5mm}   \overline{I}<\frac{xV_g}{rz}.
\end{align}

Upon A1--A2, the expression of PCC voltage is reduced to,
\begin{align}
        V(I_{\rm q})=\sqrt{V_g^2-(rI_{\rm q})^2}-xI_{\rm q}
\end{align}
which is continuously differentiable everywhere over the feasible set $\mathcal{C}$, of which the first and second order derivatives are given as,
\begin{align}\label{Vderivative}
\nabla V&=\frac{-r^2I_{\rm q}}{\sqrt{V_g^2-(rI_{\rm q})^2}}-x\\
\nabla^2 V&=\frac{-r^2V_g^2}{\left({V_g^2-(rI_{\rm q})^2}\right)^{\frac{3}{2}}}<0.
\end{align}

Thus, $V$ is strictly concave and has a unique maximum $V^\star$ over $\mathcal{C}$,
\begin{align}\label{Vmax}
    \hspace{5mm}{V}^\star=\left\{\hspace{-2mm}\begin{array}{ll}
        V(-\overline{I}), & \text{if}\hspace{2mm} {\rm C4}\hspace{2mm}\text{holds}  \\[2mm]
        V\left(\dfrac{-xV_g}{rz}\right),& \text{otherwise}.
    \end{array}
\right.
\end{align}

Based on A3, we have $V^\star<V_n$.
%If C4 holds, $V$ decreases monotonically within $[-\overline{I},\overline{I}]$; otherwise, $V$ is monotonically increasing within $[-\overline{I},-xV_g/(rz)]$ and monotonically increasing within $[-xV_g/(rz),\overline{I}]$.

Correspondingly, the first and second-order derivatives of $F$ are expressed as,
\begin{align}\label{gradF}
    \nabla F&=\eta (V-V_n)\nabla V+\frac{1}{\varepsilon}I_{\rm q}\\
    \nabla^2 F&=\eta(\nabla V)^2+\eta\left(V-V_n\right)\nabla^2V+\frac{1}{\varepsilon}.
\end{align}

Since RE minimizes a continuous function $V$ over a compact set $\mathcal{C}$, the optimum of RE  exists.

\subsection{Proof of Theorem 1}
\emph{Proof of Part 1):} 
Since C4 holds, we have $\nabla V<0$ over $\mathcal{C}$; and therefore, $\eta=\nabla V(I_{\rm q}^\ast)>0$. Then, based on A3, it follows that,
\begin{align*}
\nonumber  \nabla^2 F&=\eta(\nabla V)^2+\eta\left(V-V_n\right)\nabla^2V+\frac{1}{\varepsilon}\\
\nonumber        &>\eta(\nabla V)^2+\eta(V^\star-V_n)\nabla^2 V+\frac{1}{\varepsilon}&\\
        &>0
\end{align*}
for any $I_{\rm q}\in\mathcal{C}$. Therefore, $F$ is strictly convex and RE has a unique minimizer.

For the equilibrium $I_{\rm q}^\ast$ satisfying (\ref{Equilibrium}), based on the \emph{projection theorem} \cite{BDP1999}, we have,
\begin{align*}
    \Big(\varepsilon\big(V(I_{\rm q}^\ast)-V_n\big)-I_{\rm q}^\ast\Big)\left(I_{\rm q}-I_{\rm q}^\ast\right)\leq0,\hspace{3mm}\forall I_{\rm q}\in\mathcal{C}
\end{align*}
which yields [c.f. (\ref{gradF})],
\begin{align*}
    \nabla F(I_{\rm q}^\ast)\left(I_{\rm q}-I_{\rm q}^\ast\right)\geq0,\hspace{3mm}\forall I_{\rm q}\in\mathcal{C}.
\end{align*}

Since RE is a convex, it can be concluded that $I_{\rm q}^\ast$ is exactly the unique optimum of RE. This completes the proof of the first part of Theorem 1. \qed

\emph{Proof of Part 2):}  Now, we prove the second part.

According to (\ref{Vmax}), $I_{\rm q}=-\overline{I}$ maximizes $V^\star$. Thus, the issue is equivalent to find the condition that guarantees that $\underline{I}$ is an equilibrium, which is elaborated as follows.

\begin{itemize}
    \item{\bf Necessity.} 
      Geometrically, $(-\overline{I},V(-\overline{I}))$ should be an intersection point between the  physics $V=V(I_{\rm q})$ and the droop control rule $I_{\rm q}=[\varepsilon(V-V_n)]_{\mathcal{C}}$ on the plane. When $I_{\rm q}=-\overline{I}$, it reaches the lower limit and thus belongs to the saturation regime of the piece-wise droop rule. Therefore, the saturation voltage (i.e., the knee point, e.g., $0.5$pu in Fig. \ref{drooprule}) of droop rule should be no lower than $V(-\overline{I})$, i.e.,
\begin{align*}
    \frac{-\overline{I}}{\varepsilon}+V_n\geq V(-\overline{I})
\end{align*}
which yields (\ref{SNC_droop}).
\item{\bf Sufficiency.} If (\ref{SNC_droop}) holds, we have
\begin{align*}
    \varepsilon\big(V(-\overline{I})-V_n\big)\leq-\overline{I}
\end{align*}
and therefore,
\begin{align*}
    -\overline{I}=\Big[\varepsilon\big(V(-\overline{I})-V_n\big)\Big]_{\mathcal{C}}
\end{align*}
which indicates $-\overline{I}$ is an equilibrium.  

This completes the proof of Theorem 1.\qed
\end{itemize}
\subsection{Proof of Theorem 2}
The proof of Theorem 2 is in spirit close to that of Theorem 1 but is much more complicated because C4 does not hold. 

\emph{Proof of Part 1):} Since $\nabla V(I_{\rm q}^\ast)<0$, The proof of 1) is same as the proof of Theorem 1. \qed

\emph{Proofs of Parts 2) and 3):} Similarly, based on (\ref{Equilibrium}) and the \emph{projection theorem}, we have,
\begin{align*}
    \nabla F(I_{\rm q}^\ast)\left(I_{\rm q}-I_{\rm q}^\ast\right)\geq0,\hspace{3mm}\forall I_{\rm q}\in\mathcal{C}.
\end{align*}

Since $I_{\rm q}^\ast>-\overline{I}$, we have $\nabla F(I_{\rm q}^\ast)=0$.

Therefore, if $\nabla^2 F(I_{\rm q}^\ast)>0$,  $I_{\rm q}^\ast$ is a local minimizer of $F$; if $\nabla^2 F(I_{\rm q}^\ast)<0$, $I_{\rm q}^\ast$ is a local maximizer of $F$.\qed

\emph{Proof of Part 4):}
Based on $\nabla V(I_{\rm q}^\ast)=0$ and (\ref{Vderivative})--(\ref{Vmax}), we can obtain that  $I_{\rm q}^\ast=-xV_g/(rz)$ and $V(I_{\rm q}^\ast)=V^\star=zV_g/r$. Since C4 does not hold, $I_{\rm q}^\ast$ belongs to the linear part of the droop rule, and therefore,
    \begin{align*}
   -\dfrac{xV_g}{rz}=\varepsilon\left(\frac{z}{r}V_g-V_n\right)
\end{align*}
which yields (\ref{NC_droop}).
This completes the proof. \qed

\section{Proof of Theorem 3}
It is nontrivial to solve g-ODVS with all the nonlinear constraints simultaneously. So, the basic idea of this proof will explore the exact relaxation and equivalent of g-ODVS under specific conditions, which are tractable. 

As elaborated in Section IV-B, we have known that the current and maximum active power limits are the critical ones. In this context, let us first construct several auxiliary  optimization programs as follows.
\begin{align}\label{AP0}
\notag\hspace{-5mm}{\bf ODVS}\text{-}{\bf m0}:\hspace{3mm} \underset{I_{\rm d},I_{\rm q}}{\rm min}\hspace{3mm} &-V(I_{\rm d},I_{\rm q})\\
\notag{\rm s.t.}\hspace{3mm} & I(I_{\rm d},I_{\rm q})\leq\overline{I}\\
& P(I_{\rm d},I_{\rm q})\leq\overline{P}.
\end{align}
\begin{align}\label{AP1}
\notag\hspace{-5mm}{\bf ODVS}\text{-}{\bf m1}:\hspace{3mm} \underset{I_{\rm d},I_{\rm q}}{\rm min}\hspace{3mm} &-V(I_{\rm d},I_{\rm q})\\
{\rm s.t.}\hspace{3mm} & I(I_{\rm d},I_{\rm q})\leq\overline{I}.
\end{align}
\begin{align}\label{AP2}
\notag\hspace{-5mm}{\bf ODVS}\text{-}{\bf m2}:\hspace{3mm} \underset{I_{\rm d},I_{\rm q}}{\rm min}\hspace{3mm} &-V(I_{\rm d},I_{\rm q})\\
\notag{\rm s.t.}\hspace{3mm} & I(I_{\rm d},I_{\rm q})=\overline{I}\\
& P(I_{\rm d},I_{\rm q})=\overline{P}.
\end{align}
\begin{align}\label{AP3}
\notag\hspace{-5mm}{\bf ODVS}\text{-}{\bf m3}:\hspace{3mm} \underset{I_{\rm d},I_{\rm q}}{\rm min}\hspace{3mm} &-V(I_{\rm d},I_{\rm q})\\
{\rm s.t.}\hspace{3mm} & P(I_{\rm d},I_{\rm q})=\overline{P}.
\end{align}

For brevity, the feasible sets of g-ODVS,  ODVS-m0, ODVS-m1, ODVS-m2, and ODVS-m3 are denoted as $\mathcal{F}, \mathcal{F}_0, \mathcal{F}_1, \mathcal{F}_2$ and $\mathcal{F}_3$, respectively. The optima of g-ODVS,  ODVS-m0, ODVS-m1, ODVS-m2, and ODVS-m3 are defined as
$u^\star, u^\star_0, u_1^\star, u_2^\star$ and $u_3^\star$, respectively; correspondingly, the resultant maximum voltage are defined as $V^\star, V^\star_0, V_1^\star, V_2^\star$ and $V_3^\star$, respectively.

First of all, the existence of optimum is claimed as follows. 
\begin{itemize}
    \item It starts with g-ODVS. $\mathcal{F}\neq\emptyset$ since $I_{\rm d}=I_{\rm q}=0$ is obviously a feasible solution. Now, we show that $\mathcal{F}$ is \emph{compact}. $\mathcal{F}$ is the intersection of the level sets of several continuous functions, which are all closed sets. Therefore, $\mathcal{F}$ is closed. $\mathcal{F}$ is also bounded because the current constraint is bounded. Therefore, $\mathcal{F}$ is compact. Since $V$ is continuous, by the \emph{extreme value theorem}, it is {bounded}, and achieves its global minimum and maximum on $\mathcal{F}$.
    \item 
The proofs regarding ODVS-m0, ODVS-m1 and ODVS-m2 are similar to that regarding g-ODVS and  omitted for brevity. 
\item For ODVS-m3, $\mathcal{F}
_3$ is  closed. Now, we show that it is bounded. Suppose $I_{\rm d}\rightarrow\pm\infty$, by (\ref{stabilityconstr}), we have $I_{\rm q}\rightarrow\mp\infty$ and $V\rightarrow\pm\infty$. This contradicts the constraint $P=\overline{P}$. Similar results can be obtained if $I_{\rm q}$ is supposed to be unbounded. This implies that both $I_{\rm d}$ and $I_{\rm q}$ are both bounded. Therefore, $\mathcal{F}_3$ is compact. Since $V$ is continuous, it is bounded on $\mathcal{F}_3$ and achieves its global optimum. 
\end{itemize}

Observe (\ref{AP0})--(\ref{AP3}), the following facts about the relationship among the optimization programs are revealed:
\begin{description}
       \item[F1] ODVS-m0 is a relaxation of g-ODVS.
           \item[F2] ODVS-m1 is a (convex) relaxation of ODVS-m0.
        \item[F3] ODVS-m0 is a relaxation of ODVS-m2; ODVS-m0 is exact if $P(u_0^\star)=\overline{P}$ and $I(u_0^\star)=\overline{I}$. 
    \item[F4] ODVS-m3 provides a lower bound for ODVS-m0 if  $P(u_0^\star)=\overline{P}$.
\end{description}
  
The proof of Theorem 3 will be based on F1--F4. Note that function $V$ is nondifferentiable at the stability boundary $S=0$ and the programs may be nonconvex. Thus, all the KKT points, potential irregular solutions and nondifferentiable points should be checked to guarantee the global optimality. 

For better following the proof, the first-order derivatives of $V$ and $P$ are given as follows:
\begin{align*}
&\left\{\hspace{-2mm}
\begin{array}{l}\dfrac{\partial V}{\partial I_{\rm d}}=
r-\dfrac{r(rI_{\rm q}+xI_{\rm d})}{\sqrt{V_g^2-(rI_{\rm q}+xI_{\rm d})^2}}\\
\dfrac{\partial V}{\partial I_{\rm q}}=-x-\dfrac{r(rI_{\rm q}+xI_{\rm d})}{\sqrt{V_g^2-(rI_{\rm q}+xI_{\rm d})^2}}
\end{array}\right.\\
&\left\{\hspace{-2mm}
\begin{array}{l}\dfrac{\partial P}{\partial I_{\rm d}}=\dfrac{3}{2}\Bigg(
2rI_{\rm d}-xI_{\rm q}-\dfrac{V_g^2-r^2I_{\rm q}^2-2x^2I_{\rm d}^2-3rxI_{\rm d}I_{\rm q}}{\sqrt{V_g^2-(rI_{\rm q}+xI_{\rm d})^2}}\Bigg)\\
\dfrac{\partial P}{\partial I_{\rm q}}=\dfrac{3}{2}\left(-xI_{\rm d}-\dfrac{rI_{\rm d}(rI_{\rm q}+xI_{\rm d})}{\sqrt{V_g^2-(rI_{\rm q}+xI_{\rm d})^2}}\right).
\end{array}\right.
\end{align*}

Part 1) of Theorem 3 follows Lemmas 1 and 2.

\emph{Lemma 1:} ODVS-m1 has a unique optimum:
\begin{align}
    u^\star_1=\left(\frac{r}{z}\overline{I},-\frac{x}{z}\overline{I}\right)^T.
\end{align}

\emph{Proof:} We first solve the KKT points and then compare them to the irregular points and nondifferentiable points.

{\bf KKT.} The Lagrangian associated with ODVS-m1 is:
\begin{align*}
L_1(I_{\rm d},I_{\rm q},\lambda):=&-V(I_{\rm d},I_{\rm q})+\lambda\left(I(I_{\rm d},I_{\rm q})^2-\overline{I}^2\right) %
\end{align*}
where $\lambda\geq0$ is the Lagrange multiplier.

The (first-order) KKT  necessary conditions associated with ODVS-m1 reads,
\begin{subequations}\label{KKTP1}
\begin{align}
 -r+\dfrac{x(rI_{\rm q}+xI_{\rm d})}{\sqrt{V_g^2-(rI_{\rm q}+xI_{\rm d})^2}}+2\lambda I_{\rm d}&=0\\[0.5mm]
   x+\dfrac{r(rI_{\rm q}+xI_{\rm d})}{\sqrt{V_g^2-(RI_{\rm q}+xI_{\rm d})^2}}+2\lambda I_{\rm q}&=0\\
       I_{\rm d}^2+I_{\rm q}^2-\overline{I}^2&\leq0\\[0mm]
    \lambda&\geq0\\[0mm]        
    \lambda\left(I_{\rm d}^2+I_{\rm q}^2-\overline{I}^2\right)&=0
\end{align}
\end{subequations}

To derive the closed-form solution of (\ref{KKTP1}), i.e., the KKT points, two cases will be discussed: i) $\lambda=0$ and ii) $\lambda>0$.

\emph {i).} If $\lambda=0$, then from (\ref{KKTP1}a) and (\ref{KKTP1}b), we have 
\begin{align*}
    r^2=\dfrac{rx(rI_{\rm q}+xI_{\rm d})}{\sqrt{V_g^2-(rI_{\rm q}+xI_{\rm d})^2}}=-x^2
\end{align*}
which holds if and only if $r=x=0$. This contradicts the assumption in Section II.

\emph {ii).} If $\lambda>0$, $I_{\rm d}^2+I_{\rm q}^2-\overline{I}^2=0$. Multiplying (\ref{KKTP1}a) and (\ref{KKTP1}b) with $I_{\rm q}$ and $I_{\rm d}$, respectively, it follows that
\begin{align*}
    rI_{\rm q}+xI_{\rm d}=\dfrac{(xI_{\rm q}-rI_{\rm d})(rI_{\rm q}+xI_{\rm d})}{\sqrt{V_g^2-(rI_{\rm q}+xI_{\rm d})^2}}.
\end{align*}

It is again split into two sub-cases:
\begin{align}\label{KKTu1_ii''}
\notag  ii^\prime:\hspace{3mm}  
  &rI_{\rm q}+xI_{\rm d}\neq0 \Longrightarrow V_g\equiv\overline{I}z;\\
  ii^{\prime\prime}:\hspace{3mm} &  rI_{\rm q}+xI_{\rm d}=0  \Longrightarrow 
I_{\rm d}=\dfrac{r}{z}\overline{I},
I_{\rm q}=-\dfrac{x}{z}\overline{I}.
\end{align}

Therefore, the unique KKT solution is 
\begin{align}\label{KKT_u1}
   u^{\rm KKT}_1= \left(\dfrac{r}{z}\overline{I},-\dfrac{x}{z}\overline{I}\right)^T.
\end{align}

Geometrically, $u^{\rm KKT}_1$ is the intersection point between the straight $\mathcal{L}_1$ and current limit boundary $I(I_{\rm d},I_{\rm q})=\overline{I}$ in the fourth quadrant.

{\bf Irregular solution.} The unique solution of $\nabla I(I_{\rm d},I_{\rm q})=0$ is  $(0,0)^T$; however, 
$I(0,0)\neq\overline{I}$. Therefore, there is no irregular solution.

{\bf Nondifferentiable solution.} Now, we discuss the nondifferentiable points on $\mathcal{F}_1$. The solution may exist if and only if $V_g<\overline{I}z$. If $V_g\geq\overline{I}z$,  $V$ is differentiable everywhere on $\mathcal{F}_1$.

If the nondifferentiable solution exists, it should satisfy
\begin{subequations}\label{nondifferentiablepoints}
\begin{align}
{i:}\hspace{5mm}&rI_{\rm q}+xI_{\rm d}-V_g=0,\,\,{\rm or}\\
{ii:}\hspace{5mm}&rI_{\rm q}+xI_{\rm d}+V_g=0.
\end{align}
\end{subequations}

Then, by eliminating $I_{\rm q}$, $V$ is reduced to,
\begin{subequations}\label{vnondiff}
\begin{align}
 {i:}\hspace{3mm}V(I_{\rm d})&=\left(r+\frac{x^2}{r}\right)I_{\rm d}-\frac{xV_g}{r}\\
 {ii:}\hspace{3mm}V(I_{\rm d})&=\left(r+\frac{x^2}{r}\right)I_{\rm d}+\frac{xV_g}{r}.
\end{align}
\end{subequations}

Based on the currrent constraint and (\ref{nondifferentiablepoints}), we have,
\begin{subequations}
\begin{align*}
 {i:}\hspace{3mm}z^2I_{\rm d}^2-2xV_gI_{\rm d}+V_g^2-r^2\overline{I}^2&\leq0\\
 {ii:}\hspace{3mm}z^2I_{\rm d}^2+2xV_gI_{\rm d}+V_g^2-r^2\overline{I}^2&\leq0
\end{align*}
\end{subequations}
of which the solution sets are 
\begin{subequations}\label{solutionofnondiff}
\begin{align}
{i:}\hspace{3mm} &\left\{\hspace{-2mm}
\begin{array}{l}
 I_{\rm d}\geq\dfrac{1}{z^2}\left(xV_g-\sqrt{(xV_g)^2-z^2\left(V_g^2-r^2\overline{I}^2\right)}\right)\\[3mm]I_{\rm d}\leq\dfrac{1}{z^2}\left(xV_g+\sqrt{(xV_g)^2-z^2\left(V_g^2-r^2\overline{I}^2\right)}\right)
 \end{array}
 \right.\\
 {ii:}\hspace{3mm} & \left\{\hspace{-2mm}\begin{array}{l}
 I_{\rm d}\geq\dfrac{1}{z^2}\left(-xV_g-\sqrt{(xV_g)^2-z^2\left(V_g^2-r^2\overline{I}^2\right)}\right)\\[3mm]I_{\rm d}\leq\dfrac{1}{z^2}\left(-xV_g+\sqrt{(xV_g)^2-z^2\left(V_g^2-r^2\overline{I}^2\right)}\right)
 \end{array}\right.
\end{align} 
\end{subequations}

Combining (\ref{vnondiff}) and (\ref{solutionofnondiff}), it follows that,
\begin{align*}
\nonumber i:\hspace{3mm}V&=\left(r+\frac{x^2}{r}\right)I_{\rm d}-\frac{xV_g}{r}&\\
\notag&\leq\sqrt{-V_g^2+z^2\overline{I}^2}\\
 \nonumber&<\sqrt{V_g^2+z^2\overline{I}^2}\\
 \nonumber&<\sqrt{\left(V_g+z\overline{I}\right)^2}\\
 &=V\big(u^{\rm KKT}_1\big)\\
\nonumber ii:\hspace{3mm}V&=\left(r+\frac{x^2}{r}\right)I_{\rm d}+\frac{xV_g}{r}&\\
\notag&\leq\sqrt{-V_g^2+z^2\overline{I}^2}\\
 \nonumber&<\sqrt{V_g^2+z^2\overline{I}^2}\\
 \nonumber&<\sqrt{\left(V_g+z\overline{I}\right)^2}\\
 &=V\big(u^{\rm KKT}_1\big)
\end{align*}

Therefore, the optimum of ODVS-m1 is exactly the KKT, i.e., $u_1^\star=u_1^{\rm KKT}$ and $V_1^\star=V(u^{\rm KKT}_1)=V_g+z\overline{I}$. This completes the proof.  \qed

\emph{Lemma 2:} ODVS-m1 is an exact relaxation of g-ODVS if and only if C1 holds.

\emph{Proof:} 

\emph{\bf Necessity.} If the convex relaxation is exact, i.e., $u^\star=u_1^\star$. Then, we have  $$\overline{P}\geq P(u^\star)=P(u_1^\star)=\frac{3}{2}\left(\frac{r}{z}V_g\overline{I}+r\overline{I}^2\right):=P_b.$$

\emph{\bf  Sufficiency.}
Based on F1 and F2, it follows that,
\begin{align}
   \mathcal{F}\subseteq\mathcal{F}_0\subseteq\mathcal{F}_1,\hspace{3mm}
   V_1^\star\geq V_0^\star\geq V^\star. \label{fopt}
\end{align}

Obviously, if $\overline{P}\geq P_b$, we have $u_{1}^\star\in\mathcal{F}_{0}$. So, from (\ref{fopt}), it follows that
\begin{align}
    u_{0}^\star=u_1^\star,\hspace{3mm} V^\star_0=V^\star_1.
\end{align}

Then, given that $V^\star_1>0$ and $P^\star_1>0>\underline{P}$, $u_{0}^\star=u_{1}^\star\in\mathcal{F}$. From (\ref{fopt}), it follows that,
\begin{align}
      u^\star=u_{0}^\star=u_{1}^\star,\hspace{3mm} V^\star=V^\star_{0}=V^\star_{1}.
\end{align}
This indicates ODVS-m1 is exact if C1 holds. \qed

This completes the proof of Part 1).

\emph{Lemma 3:} ODVS-m3 has a unique optimum
\begin{align}
\nonumber
 u^\star_3=\Bigg(&\frac{1}{2z}\left(-V_g+\sqrt{V_g^2+{8r\overline{P}}/{3}}\right),\\
 &\frac{-x}{2rz}\left(V_g+\sqrt{V_g^2+{8r\overline{P}}/{3}}\right)\Bigg)^T.
\end{align}

\emph{Proof:} Similarly, we start with the KKT solutions.

{\bf KKT solution.} The Lagrangian of ODVS-m3 is
\begin{align}
  L_3(I_{\rm d},I_{\rm q},\mu)=-V(I_{\rm d},I_{\rm q})+\mu\left(P(I_{\rm d},I_{\rm q})-\overline{P}\right)
\end{align}
where $\mu$ is the Lagrange multiplier. 

The KKT  conditions associated with ODVS-m2 are,
\begin{subequations}\label{KKT_m3}
\begin{align}
    \frac{3}{2}VI_{\rm d}-\overline{P}&=0\\
    -\frac{\partial V}{\partial I_{\rm d}}+\frac{3}{2}\mu\left(I_{\rm d}\frac{\partial V}{\partial I_{\rm d}}+V\right)&=0\\
    -\frac{\partial V}{\partial I_{\rm q}}+\frac{3}{2}\mu I_{\rm d}\frac{\partial V}{\partial I_{\rm q}}&=0.
\end{align}
\end{subequations}

To solve (\ref{KKT_m3}), the analysis will be split into two cases: i) ${\partial V}/{\partial I_{\rm q}}\neq0$ and ii) ${\partial V}/{\partial I_{\rm q}}=0$.

\emph{i).} Given that ${\partial V}/{\partial I_{\rm q}}\neq0$, we obtain $I_{\rm d}\neq0$ by checking (\ref{KKT_m3}c). And consequently, $\mu={2}/{(3I_{\rm d})}$. Then, combining it with (\ref{KKT_m3}b), we have $V=0$. This contradicts (\ref{KKT_m3}a) since $\overline{P}>0$. Therefore, there is no  solution in this case.

\emph{ii).} Since ${\partial V}/{\partial I_{\rm q}}=0$, we have 
\begin{align}
    x+\frac{r(rI_{\rm q}+xI_{\rm d})}{\sqrt{V_g^2-(rI_{\rm q}+xI_{\rm d})^2}}=0\Longrightarrow rI_{\rm q}+xI_{\rm d}+\frac{xV_g}{z}=0\label{partialIq0}
\end{align}

Combining (\ref{KKT_m3}a) and (\ref{partialIq0}), we have
\begin{align*}
    z^2I_{\rm d}^2+zV_gI_{\rm d}-\frac{2}{3}r\overline{P}=0.
\end{align*} 

Thus, there are two KKT solutions:
\begin{align*}
\notag   u^{\rm KKT,1}_3=\Bigg(&\frac{1}{2z}\left(-V_g+\sqrt{V_g^2+{8r\overline{P}}/{3}}\right),\\
 &\frac{-x}{2rz}\left(V_g+\sqrt{V_g^2+{8r\overline{P}}/{3}}\right)\Bigg)^T\\
\nonumber    u^{\rm KKT,2}_3=\Bigg(&\frac{1}{2z}\left(-V_g-\sqrt{V_g^2+{8r\overline{P}}/{3}}\right),\\
 &\frac{-x}{2rz}\left(V_g-\sqrt{V_g^2+{8r\overline{P}}/{3}}\right)\Bigg)^T.
\end{align*}
which are the candidates for optimum. And from (\ref{KKT_m3}a), we  
$$V\big(u^{\rm KKT,1}_3\big)>0, \hspace{3mm}V\big(u^{\rm KKT,2}_3\big)<0.$$
Therefore, $u^{\rm KKT,1}_3$ is better than $ u^{\rm KKT,2}_3$.

$u^{\rm KKT,1}_3$ is the intersection point between $\mathcal{L}_2$ and the circle $I(I_{\rm d},I_{\rm q})=\overline{I}$ in the fourth quadrant.

{\bf Irregular solution.} Based on \emph{Linear Independence Constraint Qualification} \cite{BDP1999}, it is found that the optimum cannot be \emph{irregular} because the solutions satisfying $\nabla P(I_{\rm d},I_{\rm q})=0$ are:
\begin{align}\notag
\left(0,\frac{V_g}{z}\right)^T \hspace{3mm}\text{and}\hspace{3mm}
\left(-\frac{V_g}{2z},-\frac{xV_g}{2rz}\right)^T
\end{align}
which are infeasible for ODVS-m3. 

{\bf Nondifferentiable solution.} As mentioned above, the nondifferentiable points should satisfy (\ref{stabilityconstr}). So, combining (\ref{stabilityconstr}) and (\ref{KKT_m3}a), we obtain four candidates whose $I_{\rm d}$ are:
\begin{align*}
 I_{\rm d}^{\rm ndiff,1}&=\frac{1}{2z}\left(\dfrac{ -x}{z}V_g+\sqrt{\left(\dfrac{ x}{z}V_g\right)^2+\frac{8r\overline{P}}{3}}\right)\\
  I_{\rm d}^{\rm ndiff,2}&=\frac{1}{2z}\left(\dfrac{ -x}{z}V_g-\sqrt{\left(\dfrac{ x}{z}V_g\right)^2+\frac{8r\overline{P}}{3}}\right)\\
   I_{\rm d}^{\rm ndiff,3}&=\frac{1}{2z}\left(\dfrac{ x}{z}V_g+\sqrt{\left(\dfrac{ x}{z}V_g\right)^2+\frac{8r\overline{P}}{3}}\right)\\
    I_{\rm d}^{\rm ndiff,4}&=\frac{1}{2z}\left(\dfrac{ x}{z}V_g-\sqrt{\left(\dfrac{ x}{z}V_g\right)^2+\frac{8r\overline{P}}{3}}\right).
\end{align*}

Based on (\ref{KKT_m3}), we can know that both of $I_{\rm d}^{\rm ndiff,2}$ and $I_{\rm d}^{\rm ndiff,4}$ yield  negative voltage magnitude; therefore, they are obviously not the optimum. 

$I_{\rm d}^{\rm ndiff,1}$ and $I_{\rm d}^{\rm ndiff,3}$ yield positive voltage magnitude and it follows that
\begin{align}\label{Id1Id3compare}
   \frac{1}{2z}\left(-V_g+\sqrt{V_g^2+\frac{8r\overline{P}}{3}}\right)< I_{\rm d}^{\rm ndiff,1}< I_{\rm d}^{\rm ndiff,3}.
\end{align}

\emph{Proof of (\ref{Id1Id3compare}):} Consider a function regarding $\rho$: $$C(\rho):=-\rho V_g+\sqrt{(\rho V_g)^2+\frac{8r\overline{P}}{3}}$$ with $0<\rho\leq1$. Then, we have 
$$\nabla C(\rho)=-V_g-\frac{\rho V_g}{\sqrt{(\rho V_g)^2+\frac{8r\overline{P}}{3}}}<0.$$

Thus, $C(\rho)>C(1)$ if $\rho<1$. The first inequality in (\ref{Id1Id3compare}) holds. Besides, obviously, $I_{\rm d}^{\rm ndiff,1}<I_{\rm d}^{\rm ndiff,3}$ always holds.\qed

Based on  (\ref{KKT_m3}) and (\ref{Id1Id3compare}), we can obtain that $u_3^{\rm KKT,1}$ is better than the nondifferentiable solutions. 
Therefore, we can obtain that $u^\star_3=u_3^{\rm KKT,1}$. This complicates the proof. \qed

\emph{Lemma 4:} $u^\star_3$ is also the optimum of g-ODVS if and only if the  following condition C2 (or equivalent C2$^\prime$) holds:
\begin{align}
\notag{\rm C2:}& \hspace{3mm}\overline{I}\geq \sqrt{\frac{V_g^2}{2r^2}+\dfrac{2\overline{P}}{3r}+\dfrac{(x^2-r^2)V_g}{2r^2z^2}\sqrt{V_g^2+\dfrac{8r\overline{P}}{3}}} \\
\notag{\rm C2^\prime:}&\hspace{3mm}\overline{P}\leq 
\frac{3}{8r}\Bigg(\Bigg(\frac{r^2-x^2}{z^2}V_g+2r\sqrt{\overline{I}^2-\frac{x^2V_g^2}{z^2}}\Bigg)^2-V_g^2\Bigg).
\end{align}

\emph{Proof:} Firstly, we show that C1 and C2 are mutually-exclusive. That is, if C2 holds, C1 will definitely not hold and vice versa.

Define 
\begin{align*}
I_b:=&\left\|u_3^\star\right\|_2=\sqrt{\frac{V_g^2}{2r^2}+\dfrac{2\overline{P}}{3r}+\dfrac{(x^2-r^2)V_g}{2r^2z^2}\sqrt{V_g^2+\dfrac{8r\overline{P}}{3}}}. \end{align*}

Let $\nu:=\sqrt{V_g^2+{8r\overline{P}}/{3}}$, $I_b$ can be rewritten as,
\begin{align*}
    I_b&=\sqrt{\frac{1}{4r^2}\left(\nu-\frac{x^2-r^2}{z^2}V_g\right)^2+\frac{x^2V_g^2}{z^2}}.
\end{align*}

 $\nu$ monotonically increases as $\overline{P}$ increases, $\nu>V_g$, and  $-1<(x^2-r^2)/z^2<1$. Thus, $I_b$  monotonically increases as $\overline{P}$ increases.  Thus, we have
\begin{align*}
  \sqrt{I_b^2-\frac{x^2V_g^2}{z^2}}=\frac{1}{2r}\left(\nu-\frac{x^2-r^2}{z^2}V_g\right)
\end{align*}
which yields C2$^\prime$:
\begin{align*}
    \overline{P}\leq P_b^\prime:=
\frac{3}{8r}\Bigg(\Bigg(\frac{r^2-x^2}{z^2}V_g+2r\sqrt{\overline{I}^2-\frac{x^2V_g^2}{z^2}}\Bigg)^2-V_g^2\Bigg).
\end{align*}

The boundary case of C2 is: $u_3^\star$ exactly satisfies $\left\|u_3^\star\right\|_2=\overline{I}$ and $\overline{P}=P_b^\prime$. And if $\overline{P}>P_b^\prime$,  C2 (and C2$^\prime$) does not hold. 

Then, consider the two parallel lines  $\mathcal{L}_1$ and $\mathcal{L}_2$. Let
$u^\prime:=(I_{\rm d}^{\prime},I_{\rm q}^{\prime})^T$ be the intersection between $\mathcal{L}_1$ and $I(I_{\rm d},I_{\rm q})=\overline{I}$ in the fourth quadrant and $u^{\prime\prime}:=(I_{\rm d}^{\prime\prime},I_{\rm q}^{\prime\prime})^T$ be the intersection between $\mathcal{L}_2$ and $I(I_{\rm d},I_{\rm q})=\overline{I}$ in the fourth quadrant.  $P(u^\prime)=P_b$ and $P(u^{\prime\prime})=P_b^\prime$.
Based on the geometric relationship, it follows that $I_{\rm d}^\prime>I_{\rm d}^{\prime\prime}$,  $V(u^{\prime\prime})<V(u^{\prime})$, and thus, $P(u^\prime)>P(u^{\prime\prime})$. This indicates $P_b^\prime<P_b$. 

Therefore, if C2 (or C2$^\prime$) holds, C1 will not hold, and vice versa. 

So, for ODVS-m0, if C2 holds, the maximum active power constraint is binding at the optimum. This can be easily proven using the \emph{reduction to absurdity} method, which is omitted here. In this case, ODVS-m3 provides a lower bound for ODVS-m0. 

The sufficiency and necessity of C2 (or C2$^\prime$) are shown as follows.

\emph{\bf Necessity.} If $u_0^\star=u_3^\star$, we have 
\begin{align*}
    I(u_0^\star)=I(u_3^\star)=I_b.
\end{align*}

Considering $u_3^\star$ must be feasible for ODVS-m0, $I_b\geq\overline{I}$.

\emph{\bf Sufficiency.} From C2, we can know that  $u_3^\star\in\mathcal{F}_0$. Then, based on F4, it follows that $u_0^\star=u_3^\star$.

The remaining thing is to show ODVS-m0 is an exact relaxation of g-ODVS, which is given as follows:
\begin{align*}
    P(u_0^\star)&=P(u_3^\star)=\overline{P}>0>\underline{P}\\
    V(u_0^\star)&=V(u_3^\star)=\dfrac{z}{2r}\left(V_g+\sqrt{V_g^2+\frac{8r\overline{P}}{3}}\right)>0.
\end{align*}

Therefore, ODVS-m0 (and also ODVS-m3) is an exact relaxation of g-ODVS. This completes the proof.\qed

 Part 3) of Theorem 3 follows from Lemmas 3 and 4. The remaining Part 2) follows from Lemmas 5 and 6.

\emph{Lemma 5:} If neither of C1 and C2 hold, the  nonlinear equation group
\begin{align}\label{NEG}
\left\{\hspace{-2mm}\begin{array}{l}
I(I_{\rm d},I_{\rm q})=\overline{I}\\
P(I_{\rm d},I_{\rm q})=\overline{P}
\end{array}
\right.
\end{align}
has a unique solution $(I_{\rm d},I_{\rm q})$ satisfying
\begin{align}\label{anglerange}
     -90^\circ\leq{\rm atan2}\left(\dfrac{I_{\rm q}}{\overline{I}},\dfrac{I_{\rm d}}{\overline{I}}\right)\leq{\rm atan2}\left(-\dfrac{x}{z},\dfrac{r}{z}\right)
\end{align}
 
\emph {Proof:} Based on the first equation in (\ref{NEG}), $P(I_{\rm d},I_{\rm q})$ can be re-expressed in the polar coordinates, i.e., $P=P(\varphi)$ with the constant magnitude $\overline{I}$ where $$\varphi:={\rm atan2}\left(\dfrac{I_{\rm q}}{\overline{I}},\dfrac{I_{\rm d}}{\overline{I}}\right).$$

Define ${\varphi}^\prime:={\rm atan2}\left(-{x}/{z},{r}/{z}\right)$.
Recall the proof of Lemma 4, ${\varphi}^\prime$ corresponds to $u^\prime$, which is the intersection between $\mathcal{L}_1$ and current limit boundary in the fourth quadrant. And, obviously, $P(\varphi^\prime)=P_b$.

\emph{i).}
Consider, if $x\cdot0-r\overline{I}\leq-V_g$, $P(-90^\circ)=0$ is feasible (as the second example in Fig. \ref{traj}), $P(-90^\circ)=0$ and $P(\varphi^\prime)=P_b$ are two attainable points of $P$. Given that $P$ is continuous, as per the \emph{intermediate value theorem}, there exists at least one solution $-90^\circ<\varphi<{\varphi}^\prime$, such that, $p(\varphi)=\overline{P}$ for any $\overline{P}$ satisfying $0<\overline{P}<P_b$.

\emph{ii).} If $x\cdot0-r\overline{I}>-V_g$ (as the first example in Fig. \ref{traj}), the straight line $\mathcal{L}_2$ definitely has a intersection point $u^{\prime\prime}$ with the current limit boundary, of which the angle is denoted as $\varphi^{\prime\prime}$.
Therefore, we have
$$-90^\circ<\varphi^{\prime\prime}<\varphi^{\prime}<0^\circ.$$

Recall the proof of Lemma 4, $P(\varphi^{\prime\prime})=P_b^\prime$ and  $P(\varphi^{\prime})=P_b$. So, similarly, there exists at least one solution $\varphi^{\prime\prime}<\varphi<{\varphi}^\prime$, such that, $P(\varphi)=\overline{P}$ for any $P_b^\prime<\overline{P}<P_b$. 

Combining i) and ii), we complete the proof. \qed

Lemma 5 shows the existence of the solution. The following lemma will elaborate its uniqueness and optimality.

\emph{Lemma 6:} If neither of C1 and C2 hold,
 the solution satisfying (\ref{NEG}) and (\ref{anglerange}) is exactly the unique optimum of g-ODVS.

\emph{Proof:} The proof will be divided into three steps: 1) the solution is unique and the optimum of ODVS-m2, 2) ODVS-m0 is an exact relaxation of ODVS-m2, and 3) ODVS-m0 is an exact relaxation of g-ODVS.

In Lemma 5, we have elaborated that if ${\rm max}\left\{0,P_b^\prime\right\}<\overline{P}<P_b$, the solution space is nonempty. Now, we show that the solution is unique.

Recall Section IV-A, the trajectory of $V(I_{\rm d},I_{\rm q})=U$ is the half circle over the line $rI_{\rm d}-xI_{\rm q}-U=0$, denoted as $\mathcal{V}$, with the center at $\mathcal{O}(Ur/z^2,-Ux/z^2)$ and the radius equal to $V_g/z$. And the half circle is symmetric with respect to $\mathcal{L}_1$. Consequently, as $U>0$ increases, $\mathcal{O}$ moves along $\mathcal{L}_1$ and away from $(0,0)$. 
When $\mathcal{V}$ has overlap  with the current limit boundary (intersecting or tangent), the voltage level is attainable within the current limit. The intersection points are symmetric with respect to $\mathcal{L}_1$.  

Suppose, $\varphi^\star$ and $\hat\varphi$ are two solutions satisfying (\ref{NEG}) and (\ref{anglerange}); consider, $V(\hat\varphi)>V(\varphi^\star)$. Then, it follows that  $\hat\varphi\in(\varphi^\star,2{\varphi^\prime}-\varphi^\star)$ because only the solutions in this range achieve a higher voltage due to the symmetry. And one can obtain
\begin{align}\label{cosbound}
    \cos{\varphi^\star}<\cos{\hat\varphi},\,\,\forall \varphi\in\left(\varphi^\star,2{\varphi^\prime}-\varphi^\star\right).
\end{align}

Since $P(\hat\varphi)=P(\varphi^\star)=\overline{P}$ and $V(\hat\varphi)>V(\varphi^\star)$, it should have $\cos{\hat\varphi}<\cos{\varphi^\star}$. This contradicts (\ref{cosbound}). This indicates that there is no better solution than $\varphi^\star$. Therefore, $\varphi^\star$ is unique the solution satisfying (\ref{NEG}) within the range $(-90^\circ,2{\varphi^\prime}-\varphi^\star)$ and it is obviously the optimum of ODVS-m2. 

Now, we show that ODVS-m0 is an exact relaxation of ODVS-m2. From Lemmas 2 and 4, it can be known that if C1 and C2 do not hold, $P(u_0^\star)=\overline{P}$ and $I(u_0^\star)=\overline{I}$. This can be proven by leveraging the reduction to absurdity method, which is omitted here for brevity. Then, according to F3, ODVS-m0 is exact, i.e., $u_0^\star=u_2^\star:=\left(\overline{I}\cos{\varphi^\star},\overline{I}\sin{\varphi^\star}\right)^T$.

Finally, we show that ODVS-m0 is also an exact relaxation of g-ODVS. It follows that 
\begin{align*}
    P(u_2^\star)&=\overline{P}>0>\underline{P}\\
    V(u_2^\star)&=\frac{2\overline{P}}{3\overline{I}\cos{\varphi^\star}}>0.
\end{align*}

Therefore, ODVS-m0 is also an exact relaxation of g-ODVS if neither of C1 and C2 hold, i.e.,
\begin{align*}
    u^\star=u^\star_0=u^\star_2.
\end{align*}

This completes the proof of Lemma 6.
\qed

\bibliographystyle{IEEEtran}
\bibliography{references}
\end{document}